# Filter-Bank-Enabled Leaky-Wave Antenna Array Technique for Full-Band-Locked Radar System in Stitched Frequency-Space Domain

Dongze Zheng, *Member, IEEE,* and Ke Wu, *Fellow, IEEE*

***Abstract*—Inspired by the filter-bank (FB) concept that is normally used for multi-rate signal processing, an FB-enabled array technique of leaky-wave antennas (LWAs) is proposed and studied for creating full-band-locked frequency-scanning radar (FSR) systems in a stitched frequency-space domain. This is mainly for addressing the coupling dilemma between the range and angle resolutions, which is historically and naturally inherited from a conventional FSR. First of all, the frequency-modulated continuous-wave system architecture is selected to exemplify and recall the characteristics of a conventional FSR with an emphasis on that resolution coupling. Then, a radar solution featuring a stitched frequency-space domain is introduced for the resolution decoupling, depending on an array of FB-enabled LWA channels. With the radar equation, FB-related conditions for realizing the critical "frequency-space stitching" are analytically derived and equivalently converted into several design specifications of an array of LWAs, i.e., engineered beam-scanning functions, beam-crossovers, and phase alignments. To facilitate the practical implementation of such an array technique, a detailed and generalized design flow is developed. Finally, for a simple proof of concept, an FB-enabled two-channel LWA array is modeled, fabricated, and measured. Simulated and measured results are in a reasonable agreement, and both demonstrate the desired "frequency-space stitching" behavior, i.e., enhanced spectrum bandwidth and widened radiation beamwidth. The proposed array solution may be potentially deployed for FSR systems with a full-band-locked beam illumination and a decoupled range-angle resolution, which may present a competitor against the phased array technique.**

## I. Introduction

High-performance modern radars are normally required to detect multiple targets over a large field of view (FoV) and estimate their range, angle, and other parametric information with good resolution and accuracy. This normally requires that the radar system operates with wideband signal waveforms such as linear frequency-modulated pulse (LFM) or frequency-modulated continuous-wave (FMCW), and simultaneously its antenna front-end should provide a wideband and directive beam (i.e., this directive beam is kept fixed over the wideband spectrum of the signal waveform) that can be scanned to sample the whole targeted space, i.e., wideband spatial sampling [1]-[3]. Generally speaking, there are essentially three mainstream radar solutions that are used to perform such required wideband spatial sampling (or beam-scanning), and perhaps the most representative one is the mechanical radar in which a wideband directive antenna (e.g., parabolic reflector antenna) is mechanically scanned with the aid of electric motors and rotors [3]. Another popular method is the multi-beam scheme (or called "beam-switching"), where a group of wideband directive beams pointing toward different predetermined angles may be realized by reflector-/lens-based antennas with a cluster of feeders [4]-[7], or by an array of antennas with beamforming circuits [8]-[10]. Comparatively, the phased array technique, which uses a set of phase shifters or more often true-time-delay devices (or even full-digital phased arrays using the wideband digital beamforming technique), proves to be the most flexible approach to producing a wideband scanned directive beam among the three schemes [11].

The frequency-scanning radar (FSR) is featured with a unique status in the radar world since it can behave as a trade-off solution in terms of radar detection performances and system costs/complexities compared to the mechanical, multi-beam, and phased array radars as mentioned above [3]. This is because, for example, compared to a mechanical radar, the FSR is apparently much faster in scanning a space while dispensing with those inertial devices such as electric motors and rotors. Then, compared to the multi-beam solution, it no longer suffers from a cumbersome arrangement of illumination sources or complicated beamforming circuits. Also, compared to the phased array system, it does not necessitate expensive/lossy phase shifters or true-time-delay devices together with extra phase-controlling circuits. These merits for an FSR are substantiated by its scanned antenna beam being enabled with frequency (i.e., frequency-driven beam-scanning), which, interestingly, is also an inborn and unique feature of leaky-wave antennas (LWAs) [12][13]. Given this, FSRs, or LWA-enabled

D. Zheng was with the Department of Electrical Engineering, Poly-Grames Research Center, Polytechnique Montreal, Montreal, QC H3T 1J4, Canada. He is now with the State Key Laboratory of Terahertz and Millimeter Waves (SKLTMW), City University of Hong Kong, Hong Kong SAR, People's Republic of China (dongze.zheng@cityu.edu.hk).

K. Wu is with the Poly-Grames Research Center, Polytechnique Montreal, Montreal, QC H3T 1J4, Canada (ke.wu@polymtl.ca).

radars as will be used synonymously in this paper, have been continuously being explored and developed by researchers [14]-[25]. However, note that those above benefits of FSRs compared to other radar solutions are only related to the "*beam-scanning*" functionality, and we should bear in mind that the "*wideband*" property should also be simultaneously considered for realizing a good range resolution [1]. Unfortunately, considering the fact that each coin has two sides, one not-so-well-noticeable but intuitive observation in FSRs is that there is no possibility to achieve a beam illumination on a given target or spot over a full band of interest because of the frequency-dependent beam of radar antennas (i.e., LWAs). In other words, any given target or spot can only "see" or be illuminated by a small portion of the spectrum within an allocated full band of the signal waveform. The frequency-scanned beam, as a result, brings FSRs a dilemma of coupling regarding the range and angle resolutions as will be examined later. As a side note, this tricky issue, in historic times, was in parallel with the contradiction problem between the range resolution and detection range in pulse radars which, however, has been successfully resolved by using modulated pulse waveform (e.g., LFM pulse) together with pulse compression techniques [3]. Comparatively, the coupling problem of the range-angle resolution that was normally tolerated by engineers in developing various FSRs [14]-[25], still remains to be effectively solved to date, to the best of the authors' knowledge. Perhaps this is the fundamental reason why the FSRs have historically been prevented from being popularized as the mainstream radar solution in numerous modern application scenarios that require high radar detection performances, and these areas, therefore, have been dominated for a long time by other radar solutions, especially by the phased array [3][11].

After carefully examining the frequency-space domain on which the phased array/mechanical/multi-beam radars normally operate, as shown in Fig. 1(a), and then comparing it with the counterpart that belongs to conventional FSRs as depicted on the left-hand side of Fig. 1(b), we come up with an FSR solution exhibiting a stitched frequency-space domain (or "frequency-space stitching") as sketched on the right-hand side of Fig.1, aiming to beam-illuminate any given target over a given full-band of interest and to effectively decouple the intractable range-angle resolution problem while maintaining those aforementioned virtues that come along with the frequency-enabled beam-scanning. This idea is inspired by the filter-bank (FB) concept which is normally used in the field of multi-rate signal processing [26], and it resorts to properly stitching the frequency-space domain of an array of LWA channels. To obtain the desired stitched frequency-space domain for full-band-locked radar operations, several FB-related design specifications have to be complied with in designing the array of LWAs concerning their beam-scanning functions (BSFs), beam-crossovers, and phase alignments. These FB-related conditions, which are also termed "*stitching conditions*", are analytically derived with the aid of the radar equation [3]. Furthermore, a generalized design flow is specially developed to facilitate the design of a practical FB-enabled LWA array, and this is followed by an implementation example of a two-channel LWA array for simple demonstration purposes. Notably, in this paper, the two LWA channels are directly combined in the RF analog domain by a two-way power divider/combiner. In this case, whether the critical and highly desired stitched frequency-space domain has been realized or not can be conveniently verified from the perspective of an antenna, thereby providing a convenient and direct way to examine the effectiveness and correctness of the proposed radar front-end technique. For practical uses, these LWAs would be directly connected to separate channels and the number of LWA channels will depend on the required range/angle resolutions. The uniqueness and superiority of the proposed FB-enabled LWA array technique mainly lie in the fact that the relevant FSRs can provide full-band-locked (i.e., wideband) beam-target illuminations, and possess good design freedom in terms of the range and angle resolutions (i.e., decoupled range and range resolutions). For example, we may put the angle resolution rather than the range resolution into the first priority for consideration without using plenty of Rx channels, which contrast sharply with the prevailing phased array technique [27].

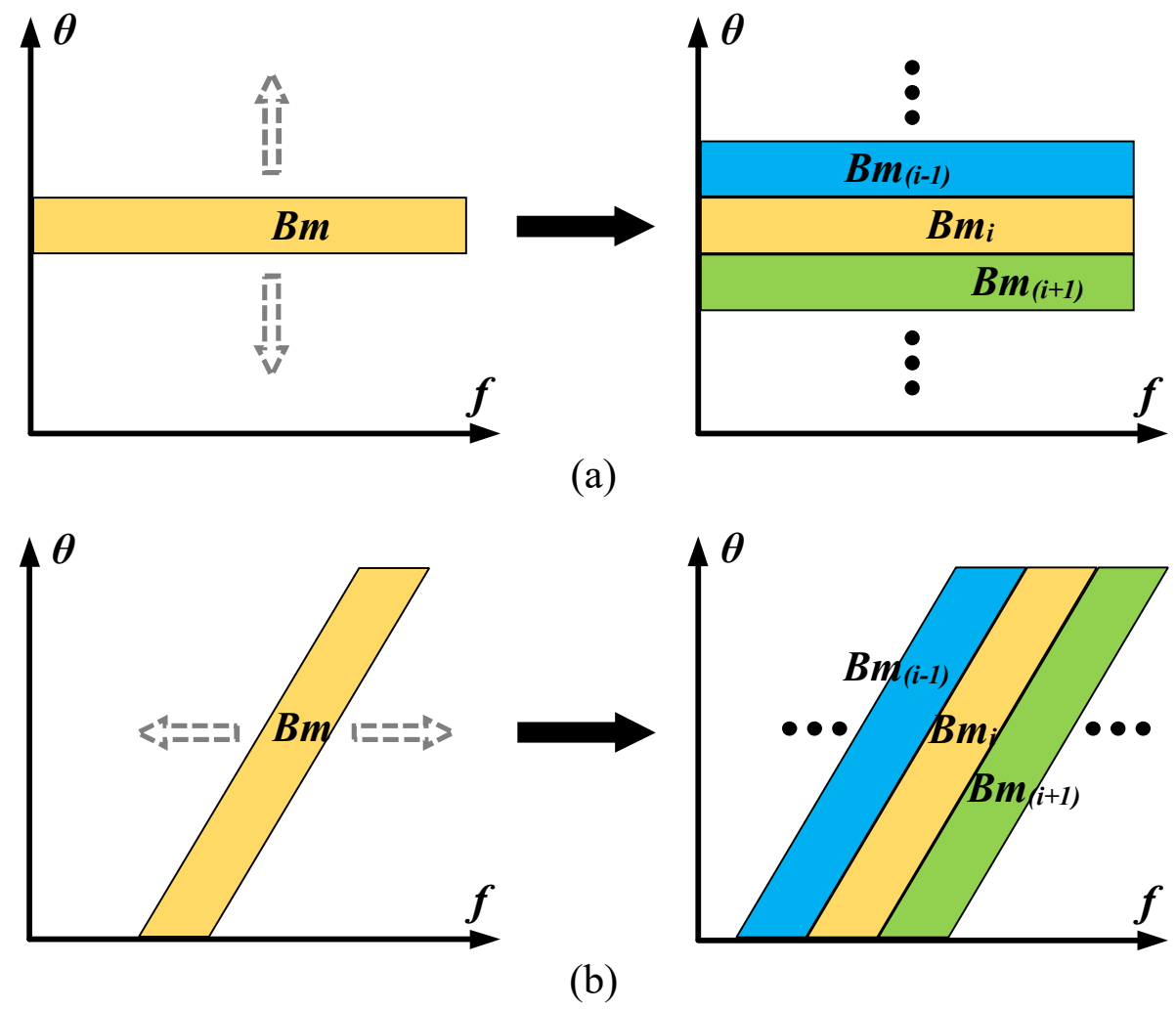

Fig. 1. Simplified characterization of radar operations with respect to different coverage plans in the frequency-space domain of signal processing [1]-[3]. (a) Phased array/mechanical/multi-beam radars. (b) LWA-enabled radars. The left-hand side of (a) corresponds to the frequency-space domain coverage of a single analog/digital beam ("Bm" in abbreviation) formed towards a specific spatial angle, while the left-hand side of (b) represents the counterpart of a single LWA channel.

## II. Background Review, Problem Analysis, and Proposed Solution

In this paper, we use the FMCW system architecture [28] as an exemplification for demonstration purposes thanks to its simplicity and popularity as well as the natural compatibility with a frequency-scanned beam. The coupling dilemma between the range and angle resolutions in a conventional FSR, as we have roughly mentioned before, will be recalled after a brief review of the operation principle of a typical FMCW radar. This is then followed by the proposed front-end solution, i.e., FB-enabled LWA array technique, to tackle that resolution coupling dilemma.

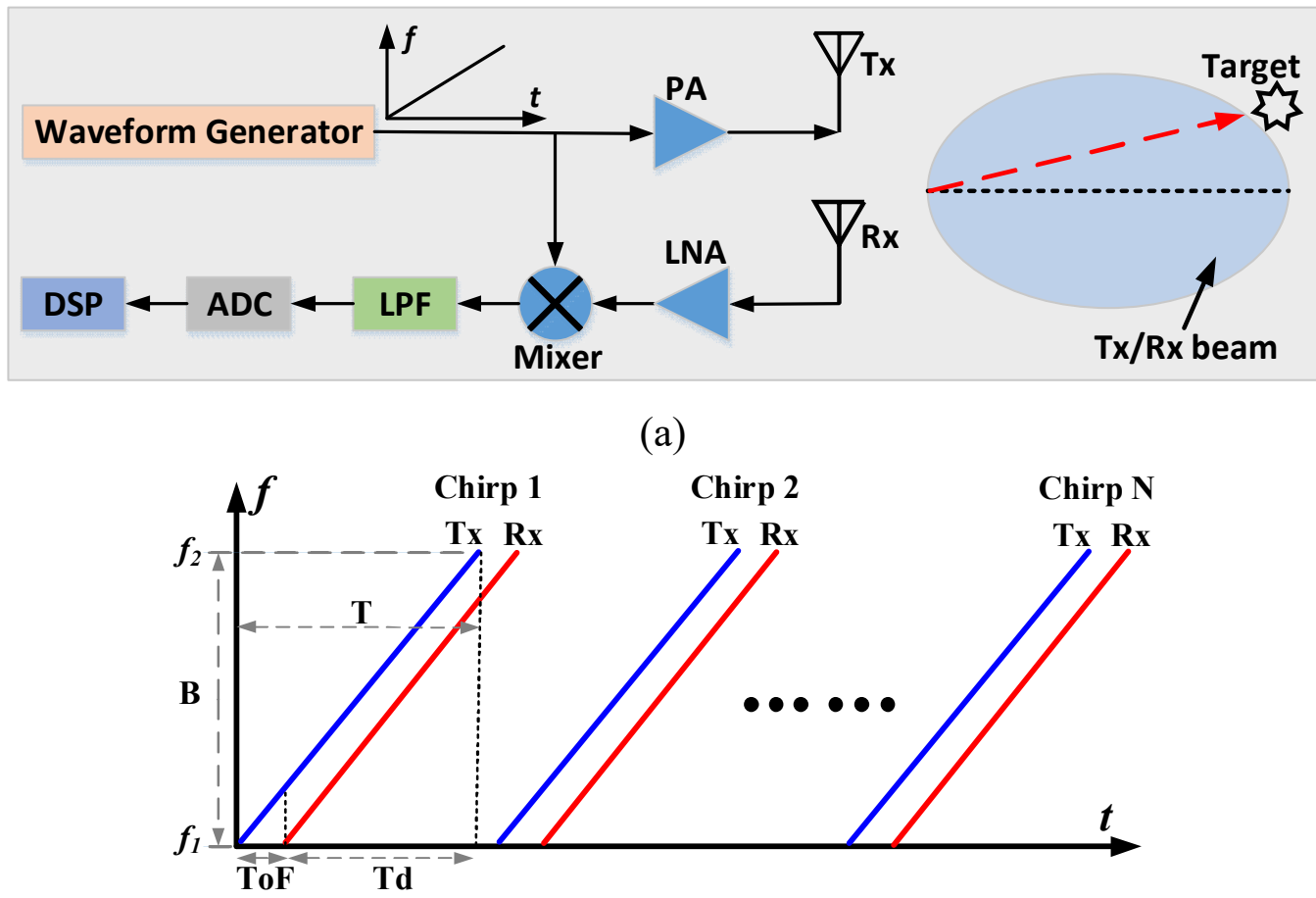

Fig. 2. (a) Simplified block diagram of a typical FMCW radar with only a single Tx-Rx chain presented. (b) Frequency-time diagram of a typical fast up-chirp-sequence waveform with $N$ chirps involved in a single frame (or CPI).

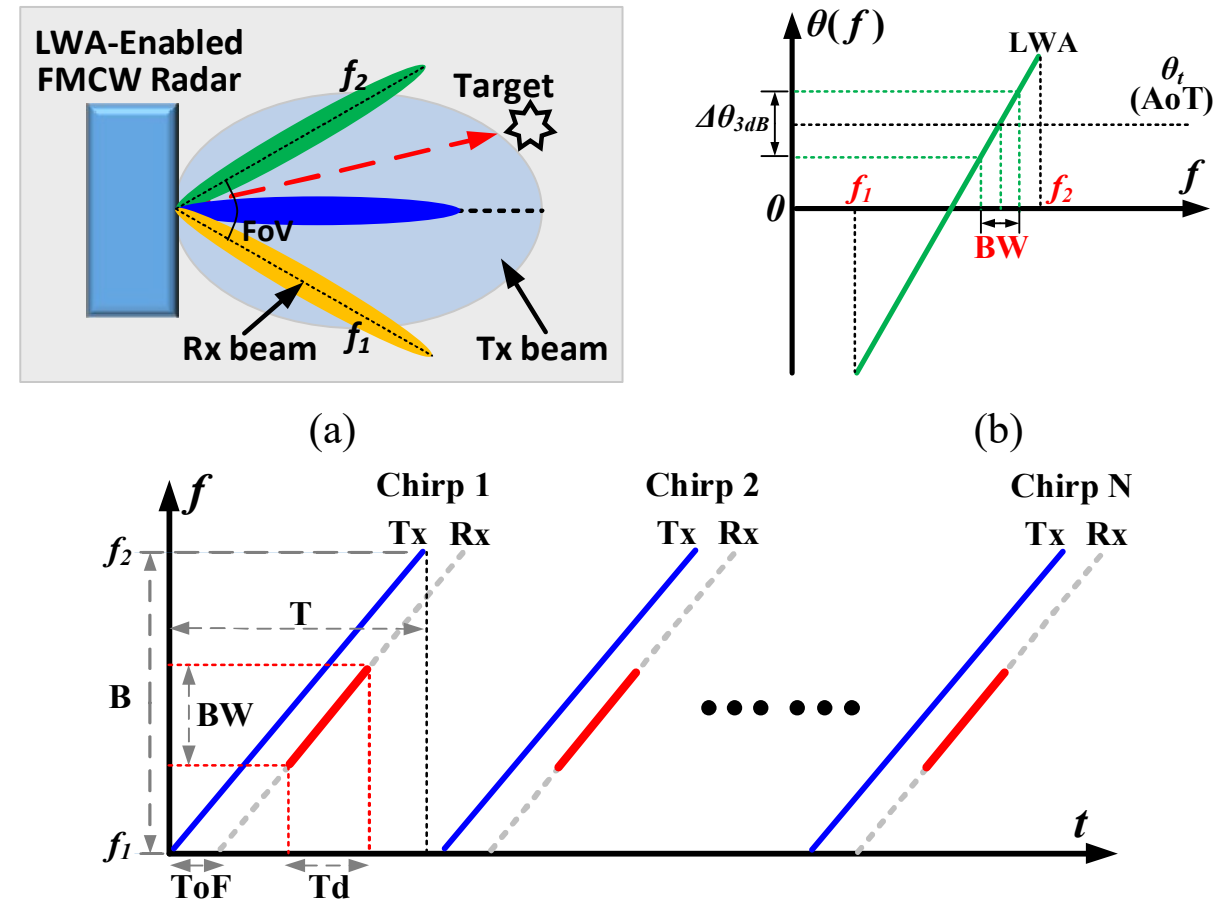

Fig. 3. (a) Simplified beam spatial coverage of an FMCW radar using an LWA as its Rx antenna, i.e., LWA-enabled FMCW radar. (b) BSF of the Rx LWA. (c) Frequency-time diagram of the transmitted and received chirp signals.

### *A. Principle of Operation and Characteristics of Typical FMCW Radar*

Fig. 2(a) shows the simplified block diagram of a typical FMCW radar system. Although in practice, like in automotive radar sensors [27][29]-[33], multiple Tx-Rx chains are normally deployed to provide good angle estimation/resolution capabilities together with improved detection performances, only a single Tx-Rx chain is presented here for illustration convenience. Fig. 2(b) depicts a popular FMCW waveform—a fast up-chirp-sequence with respect to the frequency-time diagram of transmitted and received signals within a coherent processing interval (CPI) [1][34][35]. In general, both the Tx and Rx antennas in such an FMCW radar operate with a fixed beam over the whole spectrum bandwidth of the chirp signal [36]. This means that toward any spatial directions within their beam-coverages, all the frequency components of the chirp signal would be radiated out and then captured by the radar as shown in Fig. 2(b). The relevant frequency-space domain coverage of the FMCW radar can also be described using Fig. 1(a) while the narrow beam herein is in the digital domain. The range resolution of the radar, according to [1][35], can be expressed as

$$\Delta R = \frac{c}{2B}\frac{T}{T_d} \simeq \frac{c}{2B} \qquad (1)$$

where $c$ represents the light speed in free space, while $B$ is the nominal spectrum bandwidth of the chirp signal. $T$ denotes the sweeping time of the chirp, and $T_d$ stands for the measurable time duration of the IF signal. The approximation in (1) is always made considering that the round-trip time-of-flight ($ToF$) of the signal is generally vastly smaller than the sweeping time $T$ of the chirp signal. In this sense, nearly the whole spectrum bandwidth of the chirp signal can be dedicated to the range resolution. Namely, it is the signal's spectrum bandwidth that determines the range resolution of the radar. On the other hand, multiple Rx antennas with each equipped with a channel are normally deployed to estimate and resolve the targets' angle information. A narrow beam will be produced in the digital domain and how narrow the beam mainly depends on the number of Rx channels (or virtual channels if the MIMO technique were used). The angle resolution $\Delta\theta$ of this multi-channel FMCW radar is equal to the 3-dB beamwidth $\Delta\theta_{3dB}$ of such a digital narrow beam. As for the velocity resolution, it is mainly determined by the CPI and the carrier frequency $f_c$ of the chirp signal waveform. It is irrelevant to both the available spectrum bandwidth of the received signal and the number of Rx channels, and thus it will not be involved in the following sections.

### *B. Characteristics of Conventional LWA-Enabled FMCW Radar*

Although the range resolution of a conventional multi-channel FMCW radar [27] can be easily refined by simply increasing the spectral bandwidth of the signal waveform according to (1), a good angle resolution however can only be achieved if the radar is equipped with multiple Rx antenna elements or channels. The better the angle resolution is needed, the more Rx channels should be configured in the radar. This is especially true when the "point cloud" data is needed for high-quality imaging in automotive radars for autonomous driving [37]. Apparently, this would bring several main concerns such as high cost, large system layout, heavy data volume/computation burden, complicated DSP algorithm, cumbersome channel calibration and maintenance, and severe power consumption. Consequently, only a limited number of Rx channels are deployed for most practical uses accompanied with a moderate angle resolution, e.g., 3Tx-4Rx with 12 virtual Rx channels found in the Texas Instrument (TI) product series of automotive radar sensors such as AWR1243, AWR1443, and AWR1843 [27]. In these, a good range resolution relying on the wideband FMCW signal waveform is given the first priority to guarantee reasonable radar performances, while the angle resolution only plays a secondary role.

Evolved from Fig. 2(a), Fig. 3(a) illustrates the beam-coverage scenario of a revised FMCW radar in which the Rx fixed-beam antenna is replaced by an LWA, i.e., an LWA-enabled FMCW radar. Within the working frequency band ranging from $f_1$ to $f_2$ (i.e., the bandwidth of the up-chirp waveform $B = f_2 - f_1$), the Tx beam is presumptively kept fixed [38], while the Rx beam is scanned from backward to forward as displayed in Fig. 3(b). In this sense, the Rx LWA can only receive signals when the target is "seen" by the Rx beam or when the Rx beam dwells on the target. Therefore, the spectral contents of the received signals are dependent on the spatial angles of targets, which enable the LWA to present a spatially dependent band-pass filtering behavior [39]. Here, we assume that the Rx LWA has a linear BSF $\theta_m(f)$ with a slope of $S_m$, i.e., $\theta_m(f) = S_m \cdot f + b_m$, and a constant 3-dB beamwidth $\Delta\theta_{3dB}$ that does not change with frequency. For a certain direction where the target is located, i.e., angle-of-target (AoT), the spectrum bandwidth of the received chirp signal, $BW$, is

$$BW = \frac{\Delta\theta_{3dB}}{S_m} \tag{2}$$

The spectrum bandwidth of the received chirp signal is obviously narrower than that of its transmitted counterpart since most of the frequency contents cannot be captured by the Rx LWA, as illustrated in Fig. 3(c) which exhibits the frequency-time diagram of the transmitted and received signals. Consequently, the time duration ($T_d$) of the IF signal is seriously shortened and thus the frequency resolution after performing the Range-FFT is deteriorated [1][35]. The original range resolution in (1) therefore should be modified as

$$\Delta R = \frac{c}{2BW} \tag{3}$$

which is significantly deteriorated. Considering that the angle resolution $\Delta\theta$ of the related radar is equal to the 3-dB beamwidth $\Delta\theta_{3dB}$ of the LWA, we can get

$$\Delta R \cdot \Delta\theta = \frac{c}{2} S_m \tag{4}$$

after substituting (2) into (3). Obviously, from equation (4) it is recognized that, for an LWA-enabled FMCW radar, its range and angle resolutions are coupled and contradictory for a given LWA that possesses a certain beam-scanning rate. This is easily understandable since both resolution metrics are set to rely on the 3-dB beamwidth of the LWA but have reverse trends.

### *C. Frequency-Space Domain Stitching Using Filter-Bank Concept—A Potential Solution*

Although the LWA-enabled FMCW radar, compared to the traditional counterpart shown in Fig. 2(a), can help provide a lower-cost and lower-complexity beam-scanning solution with only a single Tx-Rx chain needed in principle [14]-[25], the price it has to pay is the coupled range-angle resolution due to the nominal spectrum of the signal waveform (e.g., chirp signal) being shared by both the beam-scanning and range resolution. A straightforward observation is that a waste of the wideband signal spectrum and deterioration of the range resolution are inevitably encountered in the LWA-enabled FMCW radar, and this issue would worsen when a better angle resolution is needed according to (4). Therefore, a performance tradeoff between the attainable range and angle resolutions normally

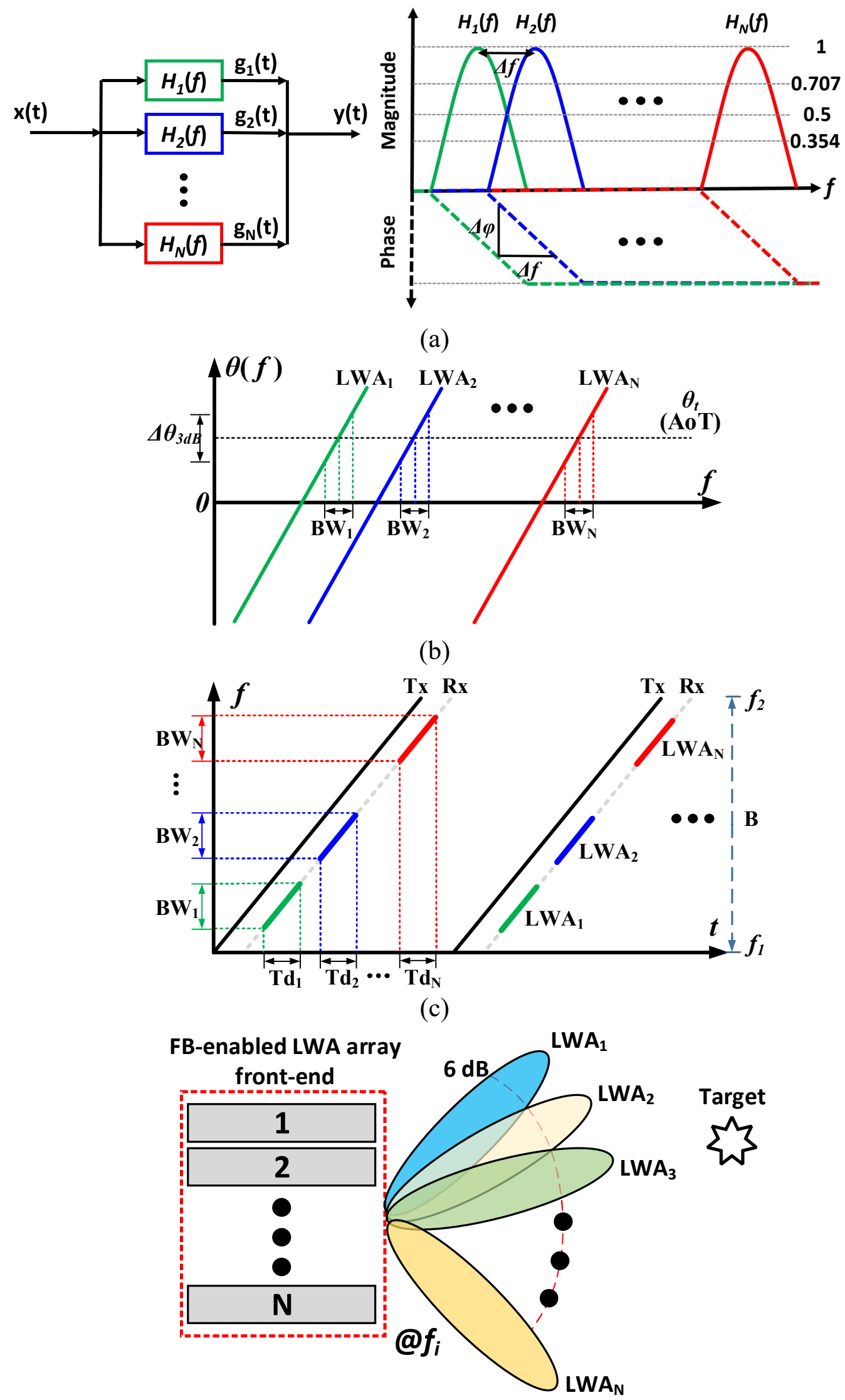

Fig. 4. Simplified illustration of an FB [26] and its associated LWA array technique for radar operations. (a) An FB with respect to its typical sub-band decomposition-summation process (left), and the ideal transfer function of each band-pass filter (right). (b) An array of FB-enabled LWAs with interleaved BSFs. (c) Frequency-time diagram of the transmitted and received chirp signals using an array of FB-based LWA channels. (d) A snapshot illustrating a possible beam configuration of the FB-enabled LWA array at a certain frequency point.

needs to be considered by radar engineers beforehand for practical uses.

As a recall, the LWA behaves as a spatially dependent band-pass filter that has different spectrum passbands toward different AoTs. Thus, for a given AoT, if two or more Rx LWA channels with elegantly interleaved passband frequency responses could be used simultaneously and their individual received spectrums could be coherently stitched in the backend module, a synthesized spectrum bandwidth wider than a single LWA could capture may be formed for radar signal processing (when understanding this design concept from the Tx perspective, a target or spot will be illuminated alternatively by the frequency-scanned beams contributed by those LWAs). For the related radar with such a multi-channel LWA array scheme, its frequency-space domain coverage would exhibit a "frequency-space stitching" phenomenon as illustrated on the right-hand side of Fig. 1(b). In this regard, the range resolution could be significantly improved according to (3), while the angle resolution would remain the same as it would with a single LWA channel. Specifically, all of these LWA channels would corporately contribute to the range resolution, while each of them may be used for the angle resolution similar to [14]-[25]. Under this context, there is one direct embodiment that the better the range resolution is needed by a system, the more LWA channels should be configured. The angle resolution, herein, is given a higher priority to be realized than the range resolution. Interestingly, this is complementary to the traditional phased-array-related multi-channel FMCW radar shown in Fig. 2 (a), where each Rx channel provides the range resolution while all of the channels work together for the angle resolution [27][29]-[35] (our later analyses will show that it is feasible for our proposed FB-enabled solution to put the range resolution in a higher priority similar to that of conventional FMCW radar, thereby suggesting that our solution still has robust design freedom in terms of the range and angle resolutions). The proposed design scheme originates from the FB concept [26] as illustrated in Fig. 4(a), where its typical sub-band decomposition-summation process and the transfer function of each band-pass filter are presented. Here, a bank of frequency-interleaved band-pass filters can be practically realized by an array of LWAs with different BSFs, as simply illustrated in Fig. 4(b). In a practical scenario, the wideband spectrum of a reflected chirp signal is first decomposed into several sub-bands by those Rx LWA channels. Subsequently, these sub-band signals are summed (in the backend DSP module) for the range resolution, while each is used for the angle resolution as mentioned. The relevant frequency-time diagram of the transmitted and received signals is conceptually plotted in Fig. 4(c), while a possible beam-configuration snapshot of the LWA array at a certain frequency point is sketched in Fig. 4(d). To obtain a gapless, stitched, and bandwidth-enhanced spectrum toward given AoTs for the related radar, as shown in the right of Fig. 1(b), the transfer function of each "band-pass filter" (i.e., LWA channel) should be deliberately organized to ensure that the summed transfer function would present a widened and relatively flat frequency response. This requirement can be translated to several stitching conditions or specifications for an array of LWAs, i.e., engineered BSFs, beam-crossovers, and phase alignments; this will be elaborately derived later.

## III. Technicality of FB-Enabled LWA Array for Frequency-Space Domain Stitching

For a single Tx-Rx chain of an FMCW radar that uses an LWA as the Rx antenna and a fixed-beam antenna as the Tx antenna, the received power spectrum, according to the radar equation [3], can be expressed as

$$P_r(\theta,f) \simeq \frac{P_t(\theta,f)G_t(\theta,f)G_r(\theta,f)\sigma\lambda_c^{\,2}}{(4\pi)^3 R^4} \tag{5}$$

where $P_r$ is the received power from the Rx antenna while $P_t$ is the transmitted power to the Tx antenna. $G_t$ and $G_r$ are the power gains of the Tx and Rx antennas, respectively. All of these parameters are a function of both frequency $f$ and spatial angle $\theta$. (Only one plane is considered here, corresponding to the scanning plane of the Rx LWA or the azimuth plane of the radar.) $R$ refers to the distance between the target and the radar; $\sigma$ represents the radar cross-section (RCS) of the target, and $\lambda_c$ denotes the free-space wavelength of the chirp signal's carrier frequency $f_c$. For the convenience of analysis without losing generality, the transmitted power $P_t$ is assumed to be a constant over frequencies, e.g., 1 watt. This is reasonable since the chirp signal waveform has a nearly uniform power spectrum [1]. Note that the Tx antenna is assumed to have a fixed beam over the frequency band of operation, and the power gain $G_t$ is thus set to be unity for simplicity—it may be different for other directions but still remain a constant over frequencies. As for the Rx LWA that has a fixed 3-dB beamwidth $\Delta\theta_{3dB}$ and linear BSF $\theta_m(f) = S_m \cdot f + b_m$, as assumed above, its absolute power gain radiation pattern is modeled by a normalized Gaussian function [2][3], i.e., Gaussian LWA, which is expressed as

$$G_r(\theta,f) = \mathrm{e}^{\left\{\frac{-2\ln 4[\theta-\theta_m(f)]^2}{\Delta\theta^2_{3dB}}\right\}} \tag{6}$$

Obviously, it can be deduced from (6) that for a certain spatial angle, the power gain frequency response of the Rx LWA is also related to a Gaussian profile. In this connection, the received power spectrum of the radar towards this angle also has an approximately Gaussian profile thanks to its linear relationship with the power gain of the Rx LWA when the above-mentioned assumptions are all considered. It should be noted that the physical significance of an antenna's transfer function is actually manifested by or equivalent to its voltage gain/field pattern [2][40]. Thus, we may simply consider

$$|H_r(\theta,f)| = \sqrt{G_r(\theta,f)} \tag{7}$$

where $H_r(\theta, f)$ represents the transfer function of the Rx LWA. The received voltage magnitude spectrum of the Rx LWA can be then expressed as

$$\begin{aligned}|U_r(\theta,f)| &= \sqrt{2\,\mathrm{Re}(Z_o)P_r(\theta,f)} \\ &= g\cdot|H_r(\theta,f)|\end{aligned} \tag{8}$$

where $Z_o$ represents the load impedance of the Rx antenna (i.e., the input impedance looking towards the receiver) while $g = \sqrt{2\,\mathrm{Re}(Z_o)\sigma\lambda_c^2 / \left[(4\pi)^3 R^4\right]}$ is a constant.

When considering the FB concept illustrated in Fig. 4(a) and employing an array of $N$ LWA channels to receive signals simultaneously, the total magnitude spectrum of the resultant voltage signal from these Rx LWA channels can be derived as

$$\begin{aligned}|U_{rt}(\theta,f)| &= \left|\sum_{j=1}^{N} U_{rj}(\theta,f)\right| \\ &= \sum_{j=1}^{N} |U_{rj}(\theta,f)|\end{aligned}$$

$$= g \cdot \sum_{j=1}^{N} \left| H_{rj}(\theta, f) \right| \tag{9}$$

where $U_{rt}(\theta, f)$ represents the spectrum of the combined voltage signal; $U_{rj}(\theta, f)$ and $H_{rj}(\theta, f)$ denote the output voltage spectrum and transfer function of the $j^{th}$ LWA, respectively. Note that all these LWAs (or "band-pass filters") here are assumed to have the same group delay (or slope of linear phase frequency responses of transfer functions) in their passbands, which is normally the case for an FB [26]. In this context, the individual output spectrums of these LWA channels could be added coherently, as they are already involved in (9), if the phase difference $\Delta\varphi$ between the adjacent LWAs in the overlap frequency region, as shown in the right of Fig. 4(a), has been compensated. Parenthetically, unless specified, it is always assumed that this phase alignment or compensation process has been automatically conducted before combining the output spectrums of those LWAs; this will be specially discussed later. As a consequence, according to (9) and Fig. 4(a), we may conclude that one key point to obtaining an enhanced spectrum bandwidth of the total received signal at a certain AoT is to make the individual transfer function of each LWA overlap at a proper frequency point. This can be translated to special design specifications of these LWAs as the transfer function of an LWA at a certain angle is determined by its BSF according to (6) and (7). In order to derive such design specifications, we consider an array of two adjacent LWA channels having interleaved BSFs expressed as $\theta_{mj}(f) = S_m \cdot f + b_{mj}$ and $\theta_{m(j+1)}(f) = S_m \cdot f + b_{m(j+1)}$. As deduced from (6) and (7), the transfer function of a Gaussian LWA, at a given angle, also has a Gaussian profile that only pertains to the frequency with the 3-, 6-, and 9-dB *spectrum bandwidth* of $\Delta\theta_{3dB}/S_m$, $\Delta\theta_{6dB}/S_m$ and $\Delta\theta_{9dB}/S_m$, respectively, where $\Delta\theta_{6dB}$ and $\Delta\theta_{9dB}$ refer to the 6- and 9-dB *radiation beamwidth* of the Gaussian LWA, respectively. To obtain a relatively flat and wide summed transfer function, their individual counterparts should approximately overlap at the frequency point where half of the maximum absolute value is reached, as illustrated in Fig. 4(a). This corresponds to the 6-dB frequency point of each transfer function, indicating that the required frequency offset or distance $\Delta f$ between the two LWAs' transfer functions is $\Delta\theta_{6dB}/S_m$. This can be translated to a limiting condition of their BSFs, expressed as

$$\begin{aligned} \theta_{mj}(f) - \theta_{m(j+1)}(f) &= b_{mj} - b_{m(j+1)} \\ &= \Delta\theta_{6dB} \\ &= \sqrt{2}\Delta\theta_{3dB} \end{aligned} \tag{10-1}$$

where a scaling factor of $\sqrt{2}$ between $\Delta\theta_{3dB}$ and $\Delta\theta_{6dB}$ of the Gaussian LWA is applied (as a side note, the scaling factor between $\Delta\theta_{3dB}$ and $\Delta\theta_{9dB}$ is $\sqrt{3}$ ). From (10-1), we can find that at a certain frequency $f$, the two adjacent LWAs should be separated by an angular spacing of $\Delta\theta_{6dB}$ with respect to their BSFs. This also means that their radiation power gain beams, at a certain frequency, should overlap at the 6-dB angular point, i.e., 6-dB beam-crossover [as sketched in Fig. 4(d)]. Equation (10-1) can thus be called the "*magnitude stitching condition*". On the other hand, recall that these design specifications related to (10-1) such as BSFs and beam-crossovers are subject to the condition that all the received spectrums from those LWA channels could be phase-aligned and thus coherently combined. However, this is not the case since there is a fixed phase difference $\Delta\varphi$ in the overlapped spectrum between the two adjacent LWAs, which can be expressed as

$$\begin{aligned} \Delta\varphi &\approx 2\pi \cdot \Delta f \cdot GD \\ &= 2\pi \cdot \frac{\Delta\theta_{6dB}}{S_m} \cdot GD \end{aligned} \tag{10-2}$$

according to the linear phase frequency responses of an FB as illustrated on the right-hand side of Fig. 4(a). Here, $GD$ refers to the shared group delay of each band-pass filter in the FB, while $\Delta f$ is selected as $\Delta\theta_{6dB}/S_m$ in accordance with the 6-dB overlap criterion adopted in (10-1). Equation (10-2) can be called "*phase stitching condition*", which should be simultaneously satisfied along with the magnitude stitching condition (10-1). If so, the frequency-space domain of each channel would be seamlessly stitched, and the individual received spectrum bandwidth from the hypothetical two Rx LWAs towards a certain angle could be coherently added and expressed as

$$BW_{FB-2} = \frac{\Delta\theta_{3dB} + \Delta\theta_{6dB}}{S_m} \tag{11}$$

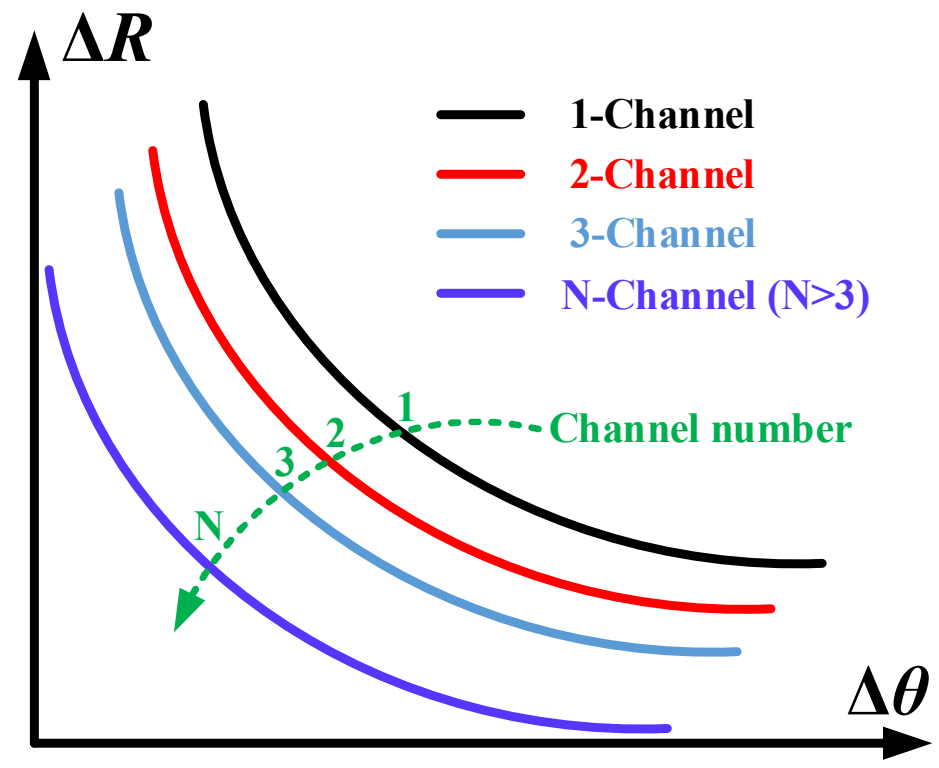

Fig. 5. Illustrative diagram based on equations (4) and (13), describing the relationship between the range and angle resolutions of a multi-channel FMCW radar using FB-enabled LWA front end.

Analogously, if an array of $N$ LWA channels is used simultaneously based on the two stitching conditions in (10), the total received spectrum bandwidth can be deduced as

$$\begin{aligned} BW_{FB-N} &= \frac{\Delta\theta_{3dB} + (N-1)\Delta\theta_{6dB}}{S_m} \\ &\simeq N\frac{\Delta\theta_{6dB}}{S_m} \\ &= \sqrt{2}N\frac{\Delta\theta_{3dB}}{S_m} \quad (N>1) \end{aligned} \tag{12}$$

As a result, the relation between the range and angle resolutions formulated in equation (4) should be revised as

$$\Delta R \cdot \Delta\theta = \frac{c}{2\sqrt{2}N} S_m \ (N > 1) \tag{13}$$

Clearly, for an FMCW radar configured with an array of $N$ Rx LWA channels that possess a certain 3-dB beamwidth (angle resolution) and beam-scanning rate, the attainable spectrum bandwidth, and range resolution could be improved by a factor of $\sqrt{2}N$ if the FB-related two stitching conditions in (10) are adopted to construct an expanded and stitched frequency-space domain for radar operations as shown in the right of Fig. 1(b). Alternatively, for a certain range resolution that has been predetermined, a refined angle resolution (narrowed 3-dB beamwidth) can be realized at the expense of the deployment of more LWA channels. An illustrative diagram describing the range-angle resolution of such a multi-channel FMCW radar using the FB-enabled LWA array technique can be found in Fig. 5. In this respect, it can be said that the proposed FB-enabled LWA array solution for FMCW radars has competitive design freedom compared to the traditional phased array counterpart in terms of the range and angle resolutions. However, if not otherwise stated, we only consider, by default, a preselected angle resolution (3-dB beamwidth) and then try to enhance the range resolution (total received signal bandwidth).

To demonstrate the effectiveness of the proposed solution for stitching the frequency-space domain, one, two, and three LWA channels of an $N$-channel ($N \geq 3$) LWA array are separately picked to calculate the total received power spectrum toward different AoTs. All of these LWAs are modeled with the same beam-scanning rates of 5°/GHz and a Gaussian radiation pattern with the 3- and 6-dB beamwidths of 8.5° and 12°, respectively. The carrier frequency $f_c$ is assumed to be 35 GHz, which is also the broadside frequency of the reference LWA numbered as $LWA_i$. It should be noted that the $LWA_i$ is the central LWA among the $N$-channel LWA array and the subscript $i$ is an integer constant that is closest to $(N+1)/2$. Based on these parameters provided above and the stitching conditions in (10), the BSFs of the three selected LWAs can be expressed as

$$\begin{cases} \theta_{m(i-1)}(f) = 5(f-35)+12 = 5f-163 \\ \theta_{mi}(f) = 5(f-35) = 5f-175 \\ \theta_{m(i+1)}(f) = 5(f-35)-12 = 5f-187 \end{cases} \tag{14}$$

The numbering and arrangement of these LWAs are consistent with Fig. 4(b). The normalized received power spectrums when using one, two, and three Rx LWAs can thus be calculated according to (6)-(9) and (14), as plotted in Fig. 6. In the case of a single LWA, it is obvious that the received power spectrum shown in Fig. 6(a) exhibits a Gaussian profile that relies on the AoT thanks to the natural spatially dependent filtering characteristics of LWAs [39]. Comparatively, when using two or three LWAs following the proposed stitching conditions in (10), spectral and spatial stitching processes, which are observed horizontally and vertically, can be accomplished as shown in Fig. 6(b) and (c), presenting an expanded coverage area of the frequency-space domain that resembles Fig. 1(b). This is because for any AoTs the received frequency components from these LWA channels would be added constructively and complementarily, thereby giving rise to a widened spectrum as compared to a single LWA. However, it is necessary to mention that the FB-related 6-dB criteria adopted in (10) are especially dedicated to obtaining a widened spectrum that particularly behaves in a relatively flat and smooth manner in the middle passband. Actually, it is not the *optimum* condition for *maximum* achievable spectrum bandwidth. Considering that the "3-dB bandwidth" is often used as a criterion in practice to define the useful spectrum bandwidth, the summed transfer function of the FB is not necessary to behave very flat whereas suitable pits or ripples in the middle of the passband can be tolerated. That is to say, the frequency distance $\Delta f$ between the adjacent transfer functions shown in Fig. 4(a) can be properly extended but not merely limited to the 6-dB criterion. For example, the overlap frequency point between two adjacent transfer functions can be relaxed to 0.354, corresponding to 9-dB stitching conditions on the analogy of (10). In this case, the summed transfer function or combined received spectrum would have pits or ripples that are approximately 3-dB lower than the peak, thereby maximizing the achievable frequency-space coverage and spectrum bandwidth in principle as illustrated in Fig. 6(d). In this respect, those equations (10)-(13) should be revised accordingly using the 9-dB-related parameters, which are not provided here for brevity. Nevertheless, it should be noted that the 6-dB associated criteria will be used by default in this paper unless otherwise stated.

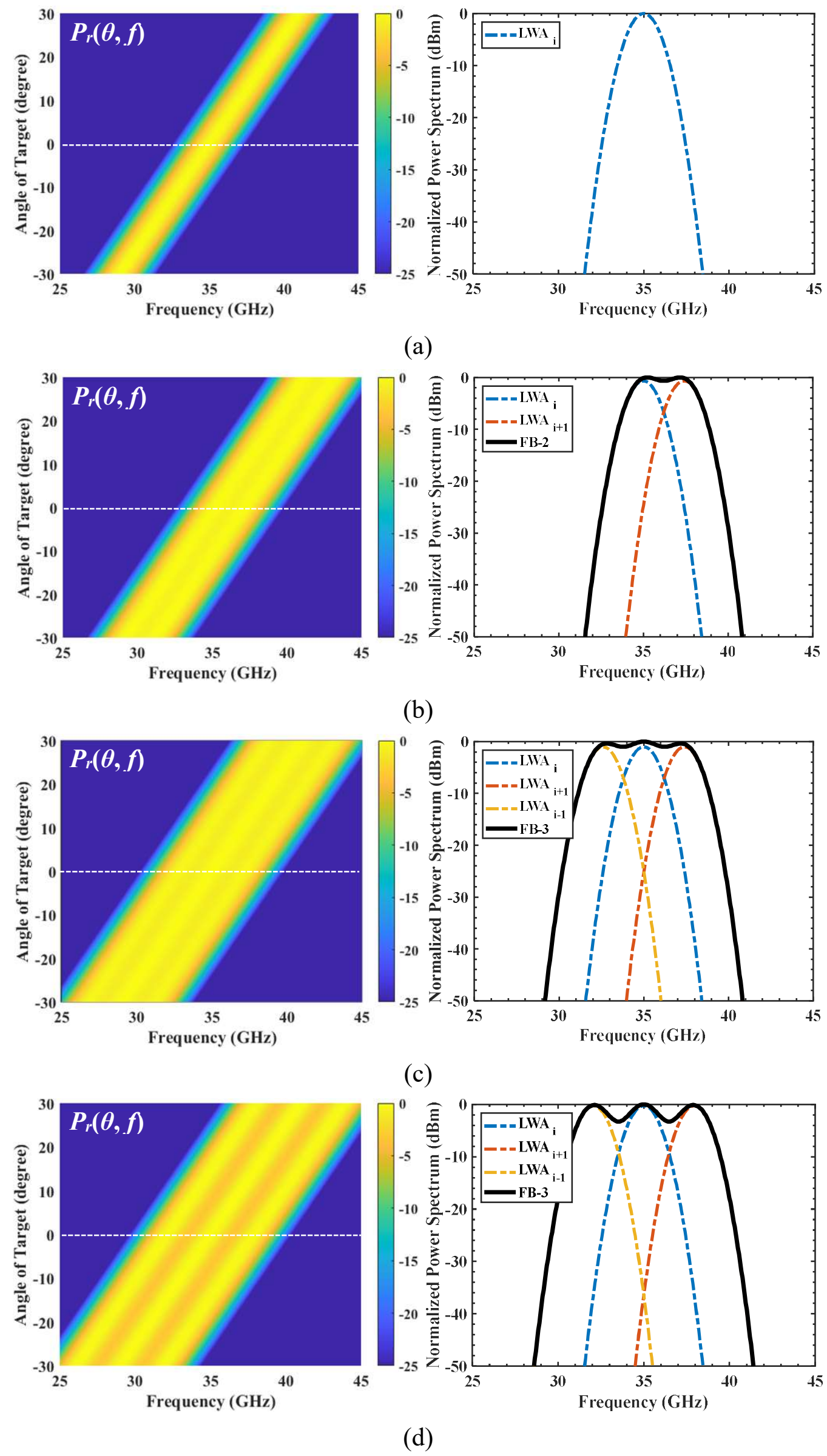

Fig. 6. Normalized received power spectrum with respect to different number of LWA channels based on the proposed FB concept. (a) A single LWA channel; (b) FB-based two LWA channels (i.e., "FB-2") under 6-dB overlap criteria; (c) FB-based three LWA channels (i.e., "FB-3") under 6-dB overlap criteria; (d) FB-based three LWA channels under 9-dB overlap criteria. The left of each subfigure represents the received power spectrum with respect to different AoTs in the manner of heat-map, while the right is related to the special situation where the target is in the broadside direction (i.e., AoT=0).

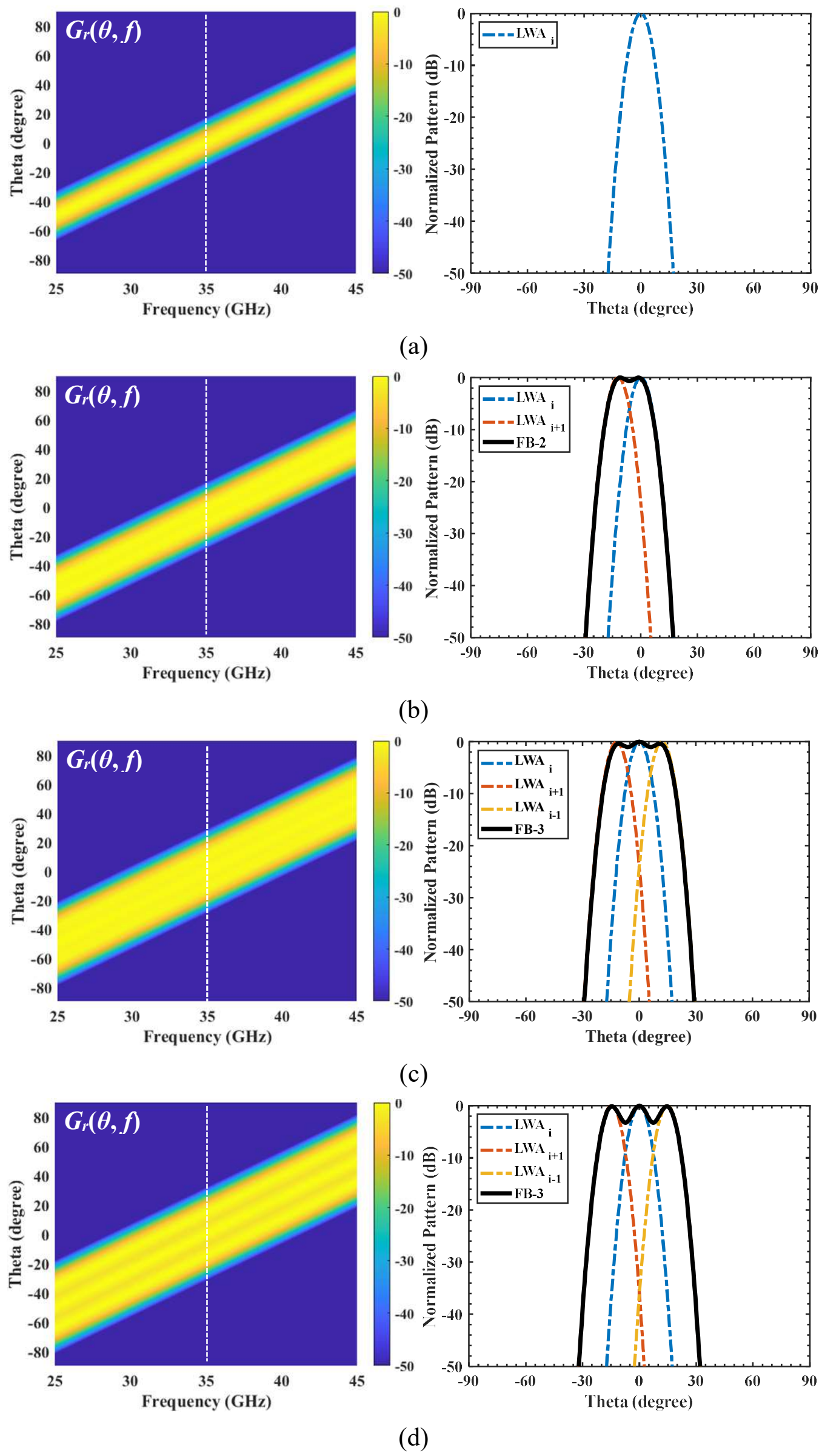

Fig. 7. Normalized radiation patterns with respect to different number of LWAs based on the proposed FB concept. (a) A single LWA channel; (b) FB-based two LWA channels under 6-dB overlap criteria; (c) FB-based three LWA channels under 6-dB overlap criteria; (d) FB-based three LWA channels under 9-dB overlap criteria. While the left of each subfigure displays the normalized radiation pattern with respect to different frequencies in the manner of heat-map, the right is related to the scenario with the carrier frequency of 35 GHz.

On the other hand, bear in mind that an antenna is a reciprocal component with respect to its Tx and Rx functionalities [40]. Examining the received power spectrum from those LWA channels to demonstrate the frequency-space stitching phenomenon in radar operations, as we have just conducted above, actually corresponds to the characterization of an antenna's Rx functionality. Nevertheless, the desirable frequency-space stitching illustrated in Fig. 1(b) can also be observed from the Tx functionality of the FB-based LWA array thanks to the reciprocal theorem; this is equivalent to the examination of the resultant radiation pattern of the whole array used as a Tx antenna. Specifically, at a certain frequency $f$, radiated signals or beams from all those LWA channels, as sketched in Fig. 4(d), will be superposed in the free space. Considering that the received power spectrum from an antenna is linearly proportional to its power gain as mentioned previously, it is expected that when examining the FB-based LWA array from the perspective of radiation, the desirable frequency-space stitching phenomenon similar to Fig. 6 would also be presented. This is unsurprisingly and clearly illustrated in Fig. 7, which depicts the normalized radiation patterns with respect to the FB-based LWA array consisting of different numbers of LWA channels that similarly follow the 6- or 9-dB stitching conditions. Those synthesized and beam-widened radiation patterns shown in Fig. 7(b)-(d) are calculated directly using the Friis transmission theorem [40], which are not expanded here for brevity. Notably, it is this widened synthesized beam that results in a prolonged beam dwell time and thus also in an enhanced received spectrum bandwidth for a target (one may also understand that the target is alternatively illuminated by each LWA channel of the whole array). Although the two characterization perspectives (i.e., Rx and Tx functionalities of the FB-enabled LWA array), from the enhanced received spectrum bandwidth at given AoTs as shown in Fig. 6 to the synthesized widened radiation beamwidth at given frequencies as shown in Fig. 7, are all valid and mutually equivalent for demonstrating the frequency-space stitching effects exhibited in Fig. 1(b), the latter is preferable and more convenient to serve as a touchstone to facilitate the practical implementation of the FB-enabled LWA array and direct proof of concept, as will be shown later.

## IV. Implementation, Simulation, and Experimentation of Practical FB-Enabled LWA Array—A Direct Proof of Concept

So far, the FB-related stitching conditions for obtaining an enlarged frequency-space coverage of an LWA-array-driven FMCW radar have been theoretically derived. Their effectiveness and correctness have been separately examined from the perspectives of the received power spectrum (Rx perspective) and radiation power pattern (Tx perspective) of the FB-based LWA front-end, where an array of LWAs with ideal electrical performances such as linear BSF, fixed beamwidth, and constant gain were assumed. In this section, we will demonstrate how to implement a practical FB-based LWA array and then provide direct proof of concept.

### A. Generalized Design Flow for FB-Enabled LWA Array

According to the leaky-wave theory [12][13], an LWA can be regarded as a special guiding-while-radiating lossy TL, where its radiation properties as an antenna can be easily predicted by its guided-wave characteristics as a lossy TL, i.e., the attenuation and phase constants. For a periodic LWA using the -1st-order space-harmonic to radiate, its main-beam direction (or BSF $\theta_m$) and 3-dB beamwidth $\Delta\theta_{3dB}$ can be expressed as

$$\theta_m(f) \simeq \sin^{-1}\left(\frac{\beta_0 - \frac{2\pi}{P}}{k_0}\right),\ \Delta\theta_{3dB} \simeq \frac{0.91}{(QP/\lambda_0)\cos(\theta_m)} \quad (15)$$

where $k_0$ and $\lambda_0$ are the free-space wavenumber and wavelength, while $P$ and $Q$ denote the period length and quantity of unit cells, respectively. $\beta_0$ refers to the phase constant of the periodic LWA's fundamental space-harmonic, which also approximates the original phase constant of the unperturbed host TL [12][13]. For a TEM or quasi-TEM TL (e.g., microstrip line) that is used for constructing a periodic LWA, the BSF in (15) can be revised as

$$\theta_m(f) \simeq \sin^{-1}\left(\sqrt{\varepsilon_{eff}} - \frac{c}{P}\frac{1}{f}\right) \quad (16)$$

where $\varepsilon_{eff}$ is the effective relative permittivity of the host TL. Thus, from (16) an advisable and feasible approach to approximately obtain an array of $N$ LWAs with engineered BSFs and beam-crossovers, which are required by the magnitude stitching condition in (10-1), is to employ a group of periodic LWAs with slightly different periods. Here, to provide an efficient design guideline and, particularly, to determine the period difference of those LWA channels, we firstly designate an LWA as the reference, which is numbered as $LWA_i$ possessing a broadside frequency of $f_{bi}$, the period length of $P_i$, and unit-cell quantity of $Q_i$. Similar to that in *Section III*, the referenced $LWA_i$ is the central one among the $N$-channel LWA array (how to determine the value of $N$ will be mentioned later), and the subscript $i$ is a known integer constant closest to $(N+1)/2$. Note that at the broadside frequency $f_{bi}$, the period $P_i$ is equal to a guided-wavelength of $LWA_i$, and the equation (16) is equal to zero, based on which $P_i$ can be calculated from $f_{bi}$ and $\varepsilon_{eff}$. Then, we assume our FoV of interest is close to the broadside direction. Consequently, in the vicinity of the broadside frequency $f_{bi}$, we may expand the $LWA_i$'s BSF at $f_{bi}$ using the Taylor Series. Keeping the first two items and neglecting higher-order ones, the BSF of $LWA_i$ can be approximately expressed as

$$\theta_{mi}(f) \simeq \frac{P_i \varepsilon_{eff}}{c} f - \sqrt{\varepsilon_{eff}} \quad (17)$$

The period difference of those LWAs is generally small and the same is true for their broadside frequencies so as to obtain well-arranged BSFs and beam crossovers. Thus, we may approximately use the referenced $LWA_i$'s BSF $\theta_{mi}(f)$, in terms of its various-order Taylor coefficients at $f_{bi}$, to expand its adjacent LWAs' BSFs $\theta_{mj}(f)$ at their individual broadside frequencies $f_{bj}$. Here, $j$ is a dummy variable that represents the numbering of those LWAs adjacent to the referenced $LWA_i$. In this case, equation (17) can be revised to give a generalized BSF suitable for all of these LWA channels (including $LWA_i$), and is expressed as

$$\theta_{mj}(f) \simeq \frac{P_i \varepsilon_{eff}}{c} f - \frac{P_i}{P_j}\sqrt{\varepsilon_{eff}} \quad (18)$$

where $P_j$ denotes the period length of the $j^{th}$ LWA, i.e., $LWA_j$. When combing (10-1) and (18) under the 6-dB design criteria, we can get

$$\begin{cases} P_j = \dfrac{\sqrt{\varepsilon_{eff}}}{(j-i)\sqrt{2}\Delta\theta_{3dB} + \sqrt{\varepsilon_{eff}}} P_i,\ j = i, i+1, i+2, \cdots\cdots \\ \text{or} \\ P_j = \dfrac{\sqrt{\varepsilon_{eff}}}{\sqrt{\varepsilon_{eff}} - (i-j)\sqrt{2}\Delta\theta_{3dB}} P_i,\ j = i-1, i-2, \cdots\cdots \end{cases} \quad (19)$$

where $\Delta\theta_{3dB}$ specifically refers to the 3-dB beamwidth of the reference $LWA_i$ at the broadside frequency $f_{bi}$. Note that (19) works only to approximately determine the initial period values of those LWAs adjacent to the reference $LWA_i$, and a fine-tuning process may be performed subsequently. The specific design flow of a generalized FB-enabled LWA array is summarized as:

1) Specify a reference periodic LWA with a certain broadside frequency $f_{bi}$ (e.g., carrier frequency $f_c$ of the chirp waveform in an FMCW radar) and a 3-dB beamwidth $\Delta\theta_{3dB}$ (i.e., angle resolution $\Delta\theta$ required by an FMCW radar). In this sense, both the period length $P_i$ and the number of unit cells $Q_i$ of this reference LWA can be approximately determined from (15) and (16). Once this reference LWA is confirmed, its beam-scanning rate $S_m$ and group delay $GD$ are also determined.
2) Based on the range resolution $\Delta R$ required by the radar, the quantity of LWA channels $N$ for implementing an FB can be estimated, according to (13) and the beam-scanning rate $S_m$ of that reference LWA. Then, the period length of the adjacent LWAs around the reference LWA can be roughly calculated according to (19) as an initial value, while the quantities of their unit cells, at this stage, can be selected approximately to that of the reference LWA.
3) Fine-tuning processes with respect to the period length and quantity of unit cells of those neighboring LWAs are necessary to ensure that all FB-related LWAs can approximately possess the designated beam-crossover and equivalent gain at the design frequency $f_c$ to coincide with (10-1), This is followed by the phase alignment procedure where the initial phase difference, according to (10-2), can be approximately calculated by considering the group delay $GD$ and beam-scanning rate $S_m$ of that reference LWA as well as the designated beam-crossover condition (e.g., 6-dB criteria).

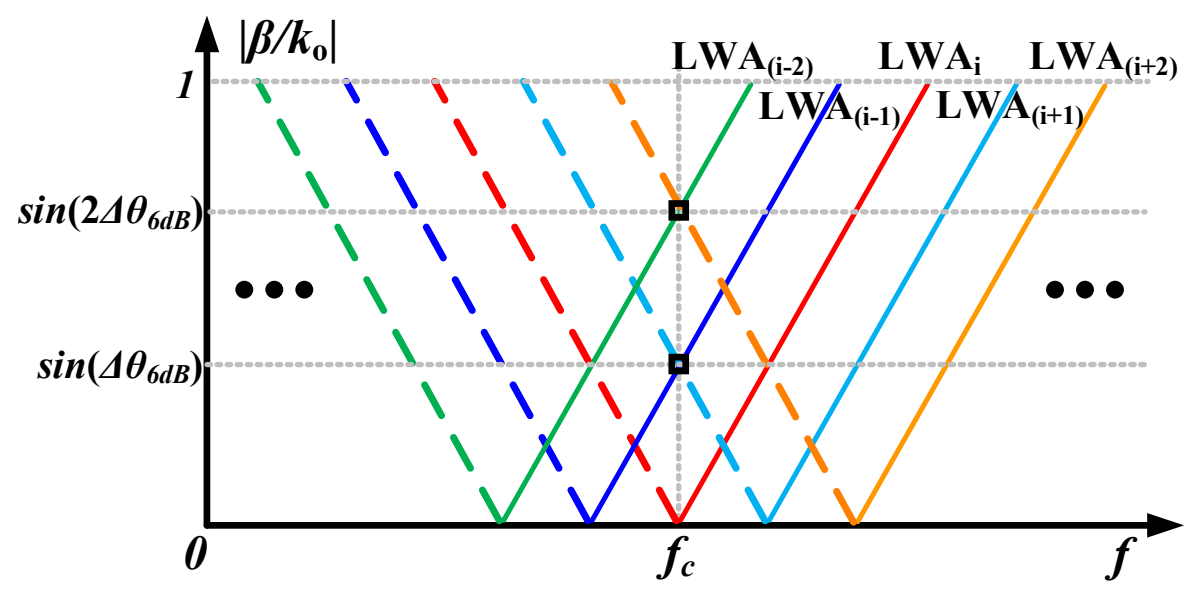

Fig. 8. Normalized phase constant of FB-enabled LWA array. $\Delta\theta_{6dB}$ refers to 6-dB beamwidth of the referenced $LWA_i$ at the broadside frequency $f_c$ (i.e., $f_{bi}$). Dashed and solid lines represent backward and forward beam-scanning regions, respectively.

Normally, for a periodic LWA, a single unit cell can be used to conveniently and accurately extract the phase constant with the help of the Bloch-Floquet theorem [41]. This step is advised since we further need to use its associated dispersion diagram as a graphical tool to facilitate the open-stopband suppression process (e.g., use some matching techniques such as the delay line and quarter-wavelength transformer [42]). As such, these FB-based periodic LWAs can support continuous beam scanning or receive signals through the broadside direction. More importantly, we can use those extracted phase constants, as conceptually illustrated in Fig. 8, to conveniently fine-tune and determine the period lengths of the adjacent LWAs (around the reference LWA) as required by the preceding three-step design flow. For example, for the referenced $LWA_i$ with a broadside frequency of $f_{bi} = f_c$ and a 6-dB beamwidth of $\Delta\theta_{6dB}$, the normalized phase constants of its two most adjacent bilateral LWAs, namely $LWA_{(i-1)}$ and $LWA_{(i+1)}$, should pass the point of [$f_c$, sin($\Delta\theta_{6dB}$)], as shown in Fig. 8. The remaining adjacent LWAs may be deduced by analogy with respect to their normalized phase constants.

### *B. Practical FB-Enabled LWA Array Using Stub-loaded-Resonator-Based Microstrip Combline LWAs*

In order to verify the proposed FB-enabled LWA array technique for the desired frequency-space stitching behavior, we will first use the 3-step design flow formulated above to implement a practical FB-based LWA array. It should be considered that the proposed FB concept illustrated in Fig. 4 needs all the transfer functions of the band-pass filters (or LWAs) to have equivalent bandwidths and maximum magnitudes. This can be technically translated to the fact that the candidate LWA should embrace stable radiation performances such as gain and beamwidth with frequency. Interestingly, recall that a class of multimode-resonator-related periodic LWAs proposed in our previous works [43][44] can tactfully satisfy such strict requirements about radiation behaviors imposed by the FB concept. As a result, the stub-loaded-resonator (SLR)-based microstrip combline LWA proposed in [44] is simply borrowed for demonstration purposes. Fig. 9 exhibits its relevant unit cell and FB-based $N$-element LWA array.

For a simple and convenient proof of concept regarding whether or not the critical "frequency-space stitching" can be achieved, we don't strictly follow the order of the 3-step design flow described above. Instead, the number of channels is determined firstly, and only two SLR-based LWAs numbered as $LWA_i$ and $LWA_{(i+1)}$ are considered here with losing generality. As already shown in Fig. 5 and formulated in (13), more LWA channels can be employed depending on the required range or angle resolution to be obtained by the relevant radar. The design frequency is selected as 35 GHz, at which the first LWA (the referenced $LWA_i$) would realize the broadside radiation. The substrate adopted here is Rogers RO3035 with a thickness of 0.508 mm, a relative permittivity of 3.6, and a loss tangent of 0.0015. The period length $P_i$ of this referenced $LWA_i$ can be determined as mentioned previously. Complete unit cell dimensions are $P_i$=5.16 mm, $W_{mi}$=0.9 mm, $W_{si}$=0.4mm, $L_{1i}$=0.6 mm, $L_{2i}$=1.2 mm, $L_{3i}$=0.3 mm, $L_{di}$=0.259 mm, $L_{qi}$=1.185 mm, and $W_{qi}$=0.422 mm. Notably, instead of specifying the 3-dB beamwidth (or angle resolution to be realized by the related radar) and then calculating the number of unit cells according to the design flow described above, we directly choose the number of unit cells of the $LWA_i$ for convenience. As a result, its 3-dB beamwidth could be determined accordingly and then used to calculate the period of the second LWA, i.e., $LWA_{(i+1)}$. Here, 10 unit cells are cascaded to form effective leaky-wave radiation for the $LWA_i$ while the remaining power is absorbed by a matching-load termination. The simulated $|S_{11}|$ and $|S_{21}|$ of this reference $LWA_i$ can be found in Fig. 10(a), showing that a good impedance matching without an open-stopband together with effective radiation is realized. The normalized radiation pattern of the $LWA_i$ at the design frequency of 35 GHz is plotted in Fig. 10(b), with the 3- and 6-dB beamwidths of 8.5° and 12°, respectively. It should be noted that for a conservative design, 6-dB-related criteria as previously mentioned are selected here. The period length of $LWA_{(i+1)}$ can be then calculated using (19), which is about 4.41 mm. Subsequently, we can use this result as an initial value to adjust the geometry of $LWA_{(i+1)}$'s unit cell by taking advantage of Fig. 8, thereby resulting in the desirable beam-crossover and the suppression of open-stopband [42]. After a fine-tuning process, the unit cell dimensions of $LWA_{(i+1)}$ are determined as $P_{(i+1)}$=4.66 mm, $W_{m(i+1)}$=0.9 mm, $W_{s(i+1)}$=0.4mm, $L_{1(i+1)}$=0.58 mm, $L_{2(i+1)}$=1.16 mm, $L_{3(i+1)}$=0.28 mm, $L_{d(i+1)}$=0.082 mm, $L_{q(i+1)}$=0.934 mm, and $W_{q(i+1)}$=0.529 mm. 11 unit cells are cascaded to establish the $LWA_{(i+1)}$ to realize an equivalent gain compared to that of $LWA_i$ at the design frequency of 35 GHz. The simulated S-parameters and normalized radiation patterns at 35 GHz of the $LWA_{(i+1)}$ are also plotted in Fig. 10(a) and (b), respectively. Notably, the desirable 6-dB beam-crossover is realized by the two LWAs at the design frequency of 35 GHz. What's more, the beam-crossover of the two LWAs is sandwiched between 6- and 9-dB over the frequency range of 32-38 GHz, as also shown in Fig. 10(b). This implies that a synthesized widened radiation beam could be realized and scanned with frequency (similar to Fig. 7), thereby paving the way for obtaining an enhanced spectrum bandwidth towards certain AoTs within the FoV (similar to Fig. 6). The simulated 2-D realized gain patterns (versus frequency and angle) of the two LWAs, together with their BSFs, are plotted in Fig. 10(c). A beam-scanning rate of about 5°/GHz and approximate gains over the frequency region of 32-40 GHz are obtained for the two LWAs. Fig. 11 exhibits the normalized received power spectrums of the two LWAs when they are illuminated by a wideband uniform plane-wave with the phase center (or zero-phase point) located at the antenna surface. When the incident angles of the plane-wave change from -10° (backward) to 10° (forward), the center frequency and passband of the received power spectrum show an increasing trend for each LWA, as expected. More importantly, the two received power spectrums approximately overlap at the designated 6-dB frequency point when the incidence angle is 0°, which is consistent with Fig. 6(b), thereby manifesting the effectiveness of the developed design flow. Also note that the spectrum crossover is still lower than 9-dB when the incident angles deviate from the broadside direction to ±10º, thereby implying that a widened spectrum with a reasonable pit or ripple could be realized within such an FoV if the two received spectrums could be added coherently. This, of course, requires a phase alignment process between the two LWA channels, as described before. As shown in the right subfigure of Fig. 11(a), the received phase spectrums of the two LWAs within their overlapped main-lobe frequency region (i.e., passband) possess a fixed phase difference of about $\Delta\varphi$≈160º, which is similar to the picture displayed in the right-hand side of Fig. 4(a). The group delay of the two LWAs in their main-lobe passbands is about $GD$≈0.198 ns, corresponding to a slope of about 71º/GHz for their phase spectrums. Considering that the reference $LWA_i$ has a beam-scanning rate of about 5º/GHz and a 6-dB beamwidth of about 12º, the required phase difference can be theoretically calculated as 170.4º according to (10-2). Both the simulated and calculated values of the phase difference are in a reasonable agreement. Notably, such a fixed phase spectrum difference between the two adjacent LWA channels can be compensated or calibrated either in the baseband digital domain by the DSP module, or in the RF analog domain by simply introducing a delay line on the second LWA (thanks to the relative narrowband property of the overlapped passband spectrum) or alternatively using the wideband phase-shifting mechanism proposed in [45]. This will be shown later. In this sense, the two output spectrums from the FB-based LWA array can be coherently superposed, thereby yielding a stitched frequency-space domain exhibiting an enhanced received spectrum bandwidth towards given AoTs within the FoV.

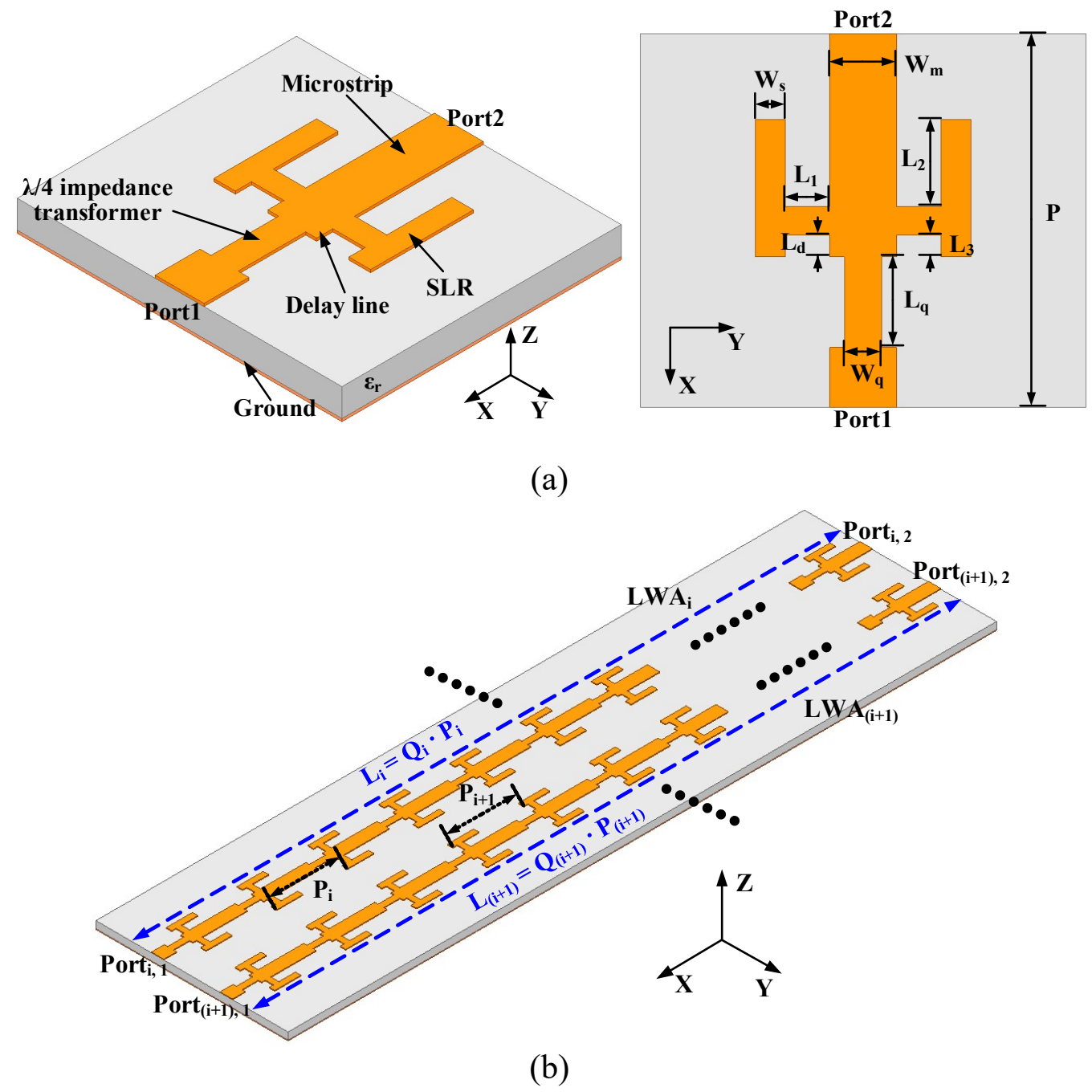

Fig. 9. SLR-based microstrip combline LWA [44] with respect to its (a) unit cell (left: 3-D view, and right: top view), and (b) FB-enabled $N$-element LWA array (only two LWA elements are exhibited for a simple illustration).

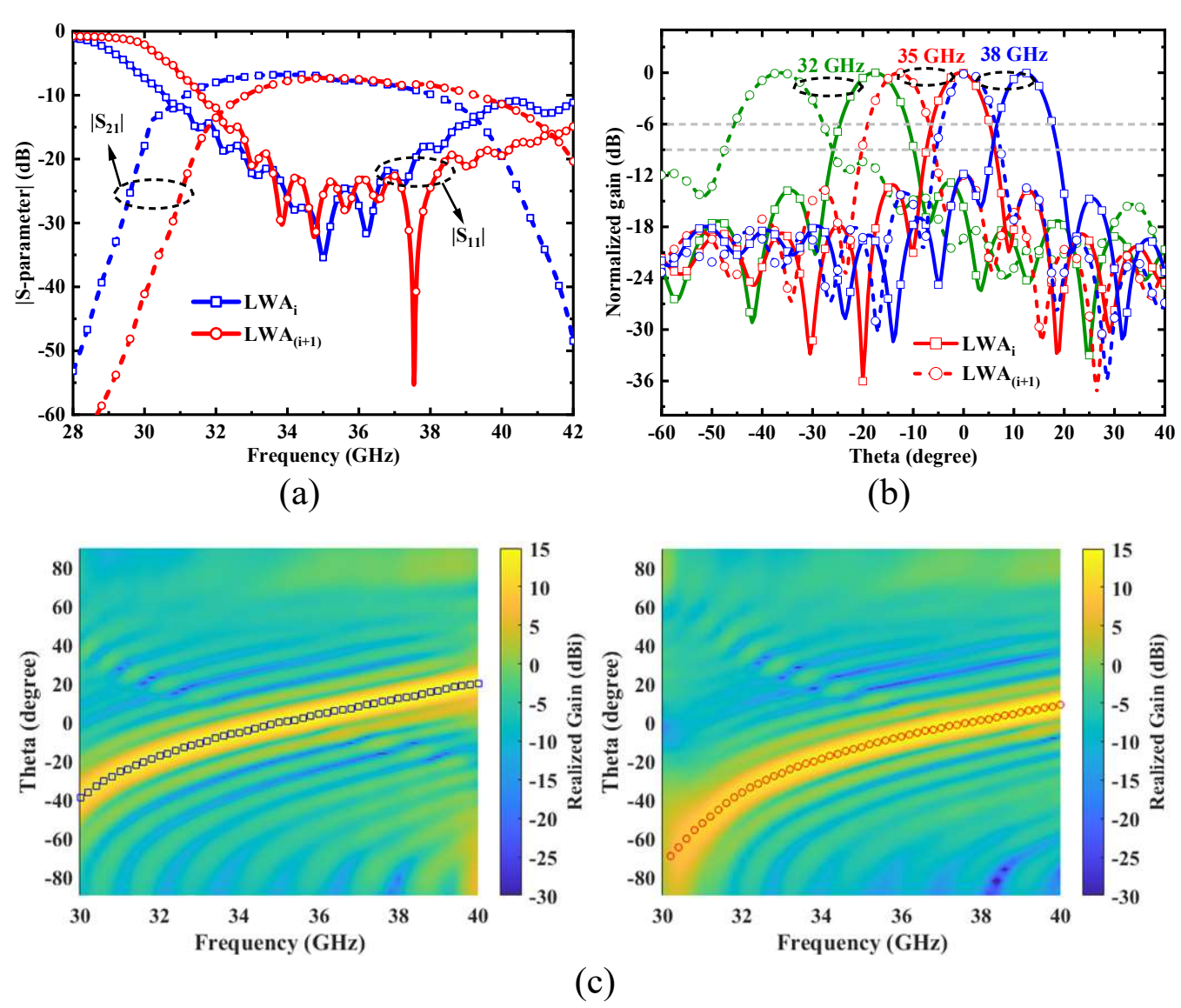

Fig. 10. Simulated electrical behaviors of the two LWAs. (a) S-parameters. (b) Normalized radiation patterns. (c) Main-beam directions and 2-D realized gain patterns (left: $LWA_i$; right: $LWA_{(i+1)}$).

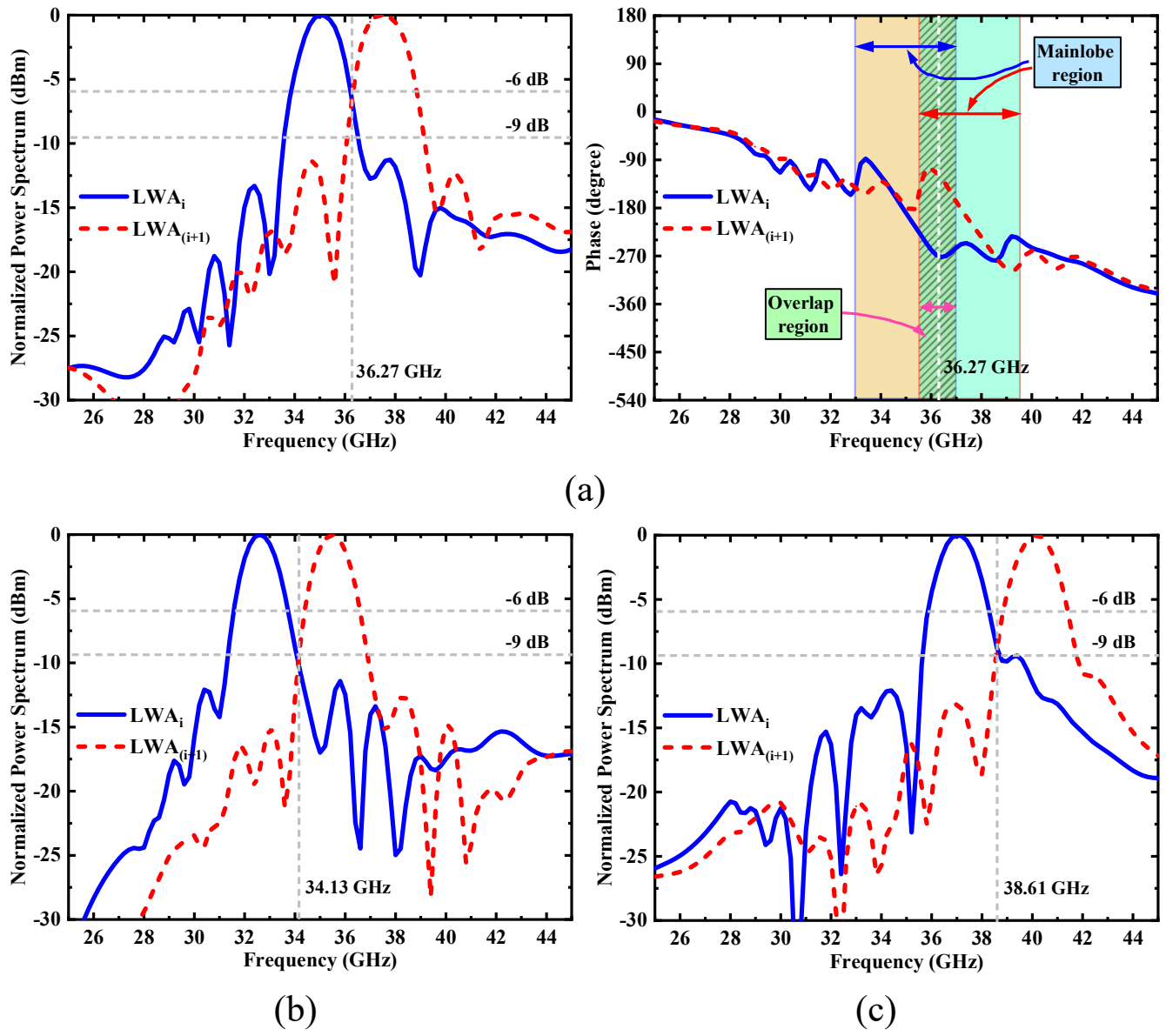

Fig. 11. Normalized received power spectrum of the two LWAs with respect to different incidence angles of a wideband uniform plane-wave. (a) 0°; (b) -10°; (c) 10°. Negative/positive angles refer to the backward/forward beam-scanning regions, respectively. In (a), the right subfigure displays the phase spectrum of received signals from the two LWAs at the broadside (0°).

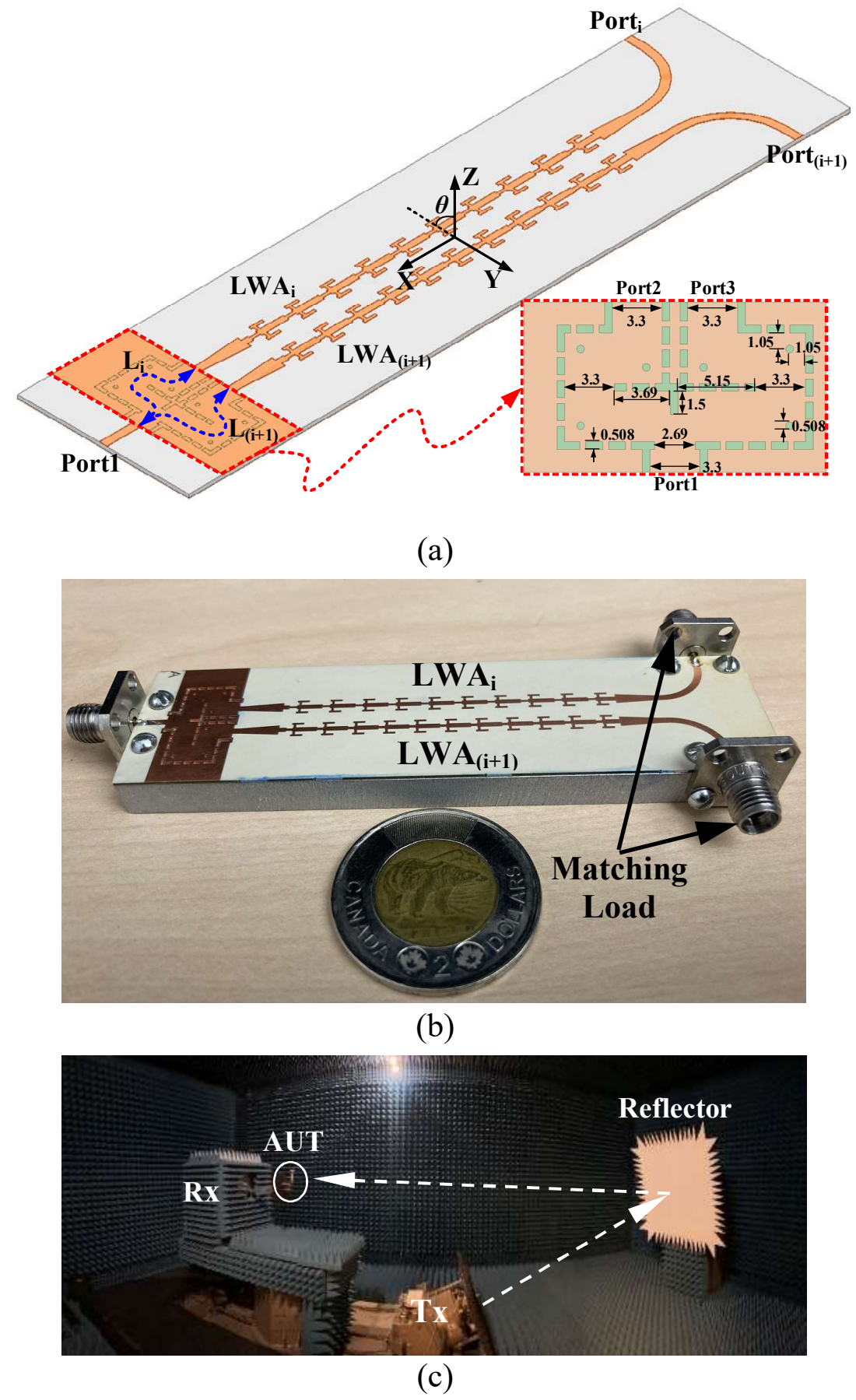

Fig. 12. Final configuration of the proposed FB-based LWA array consisting of two SLR-based microstrip combline LWAs and a two-way SIW power divider. (a) Simulated model. (b) Fabricated prototype. (c) Measurement setup in Compact Antenna Test Range. All the parameters in (a) are in millimeters.

### C. Simulation and Experimentation of FB-Enabled SLR-Based Microstrip Combline LWA Array

Our eventual goal is to collect all the received signals from those FB-related LWA channels to realize a stitched frequency-space domain coverage for radar signal processing. Thanks to the antenna being reciprocal with respect to its Rx and Tx modes, the combination process for the received signals from those $N$ LWA channels in either the digital or analog domain is equivalent to the radiation pattern synthesis of an $N$-element antenna array in the free space. Consequently, for a convenient demonstration and easy understanding of the FB-enabled LWA array technique, the antenna's Tx mode in the form of free-space signal superposition is still selected here to facilitate design and demonstration. This is just like what we have conducted to design each LWA channel in the last subsection. The proposed two SLR-based microstrip combline LWAs are connected to a two-way SIW power divider as depicted in Fig. 12(a), and the entity can be treated as a Tx two-element LWA array. Simulated S-parameters in terms of the magnitude and phase frequency responses of the power divider are plotted in Fig. 13(a) and (b), respectively, where the magnitude imbalance is less than 0.5 dB with the frequency range of 32-38 GHz. The phase difference required by the two LWA channels must be compensated so that their received spectrums (working as Rx antennas) can be coherently stitched as explained above. As such, the power divider's two branches are properly adjusted and optimized with respect to their path lengths [i.e., $L_i$ and $L_{(i+1)}$, as shown in Fig. 12(a)], where the initial path length difference can be easily determined according to the phase difference depicted in Fig. 10(a), i.e., $\Delta\varphi$≈160º. Notably, from the perspective of an antenna's Tx functionality, this phase compensation process between the two LWA channels is significant to ensure that the relative phases of their radiated fields in free space are approximately equal in the overlapped main-beam spatial region [see Fig. 4(d)]. Thus, their radiated fields can superpose coherently and form a widened radiation pattern like Fig. 7.

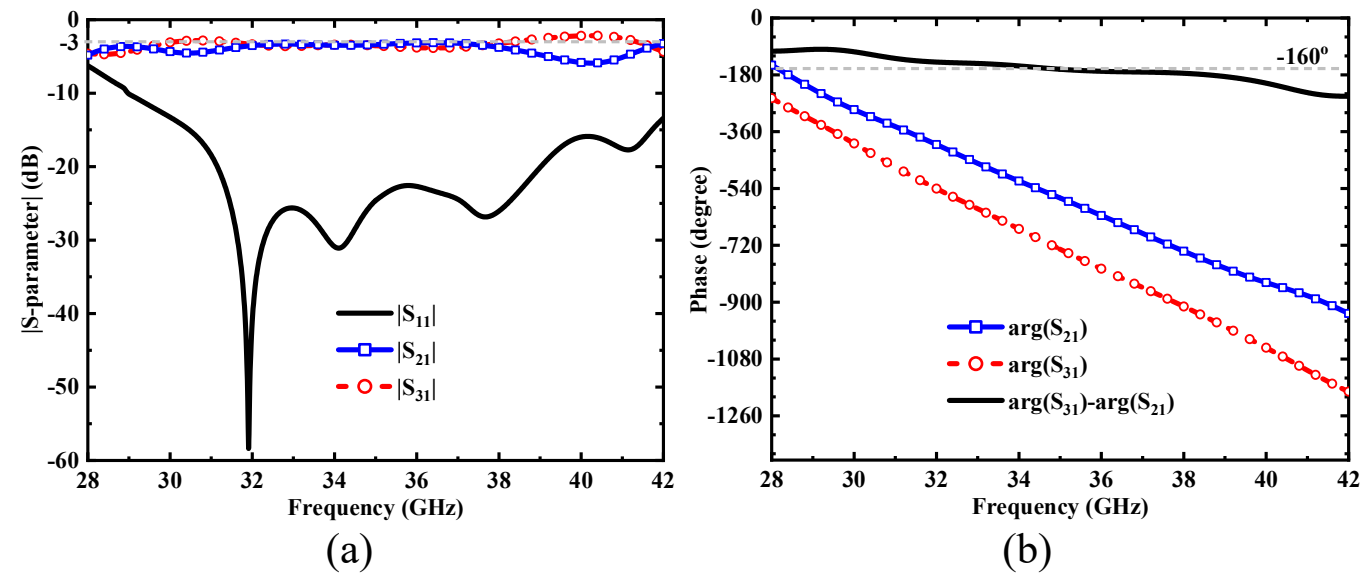

Fig. 13. Simulated S-parameters of the proposed two-way SIW power divider. (a) Magnitude. (b) Phase.

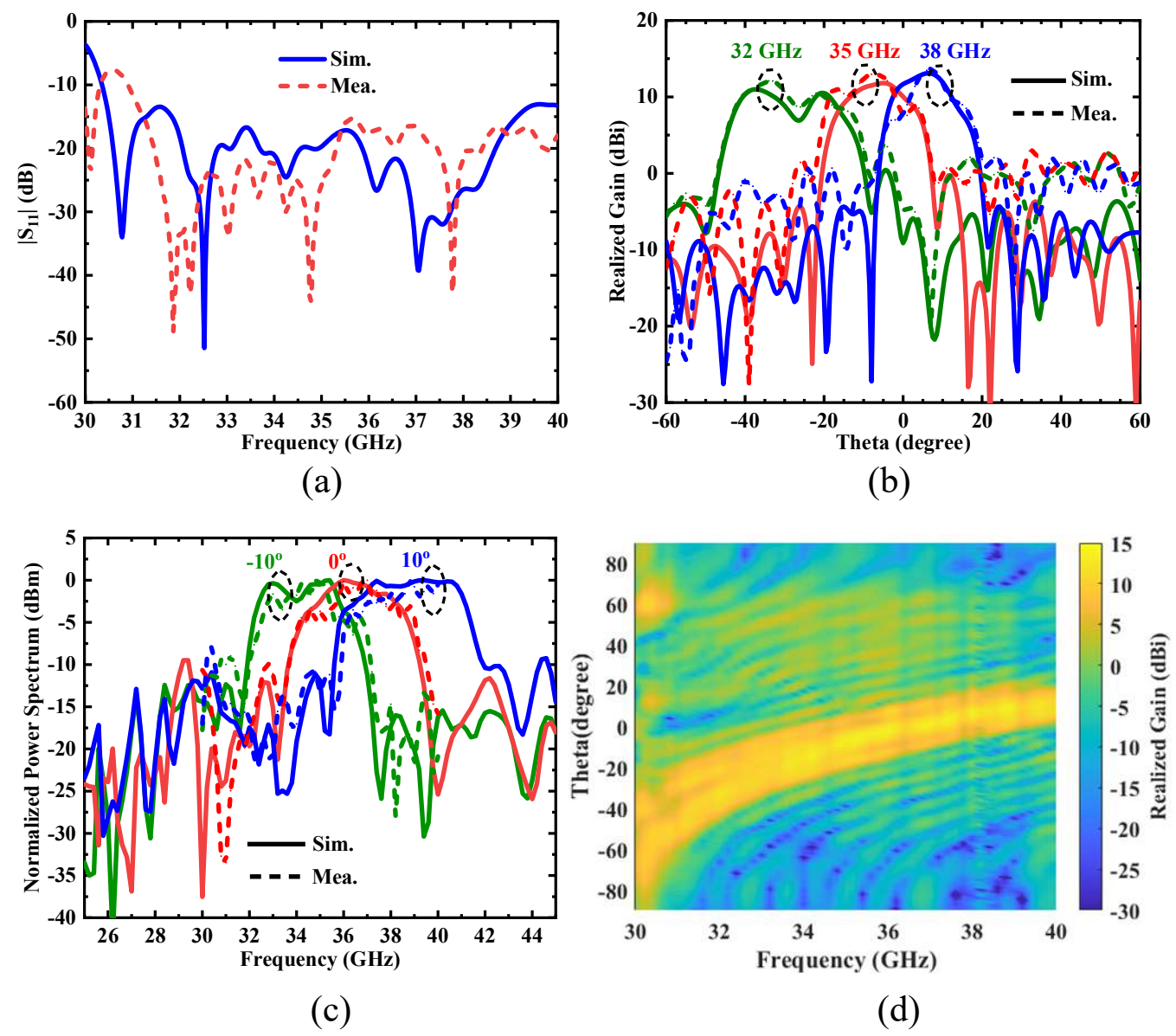

Fig. 14. Simulated and measured results of the proposed two-element FB-based LWA array. (a) $|S_{11}|$. (b) Realized gain pattern at different frequencies. (c) Normalized power spectrum towards different AoTs. (d) Realized gain as a function of both angle and frequency (only measured results). The measured results in (c) are actually the measured realized gain frequency responses.

There are three ports in the designed two-element LWA array. The common port ($Port1$) is connected to the source while the other two ($Port_i$ and $Port_{(i+1)}$) are terminated with matching loads. The simulated $|S_{11}|$ of the array can be found in Fig. 14(a), with a good impedance matching realized. Simulated radiation patterns at 32 GHz, 35 GHz, and 38 GHz are plotted in Fig. 14(b), which obviously shows that the stitched and widened radiation patterns (i.e., spatial-stitching) are realized when compared to the results of the two individual LWAs shown in Fig. 10(b). This is thanks to, first, the beam-crossover between the two LWAs being kept within a reasonable level between the 6- and 9-dB conditions even though their beams are scanned with frequency and, second, the required phase alignment process between the two LWAs being been successfully satisfied. Notably, we can then rationally speculate that the desirable spectral-stitching phenomenon and widened output spectrum would be found when the proposed FB-enabled LWA array operates in the Rx mode. To justify this, the proposed array is impinged by a wideband uniform plane-wave from different spatial angles, and simulated results in terms of the received normalized power spectrums are depicted in Fig. 14(c). As speculated, the proposed array can receive and output a stitched and approximately double-wide spectrum bandwidth compared to the situations of two individual LWAs shown in Fig. 11 and similar to the results of Fig. 6(b).

There are two aspects that should be noted before the experimental demonstration of the proposed design concept. Firstly, the measurement setup of a typical microwave anechoic chamber is always comprised of a Tx-Rx chain like Fig. 2(a), e.g., the Compact Antenna Test Range (CATR), as shown in Fig. 11(c), in which the radiation measurement of an antenna is realized by repeatedly transmitting and receiving a wideband chirp signal against different spatial angles [40]. This is what we need for our experimentation. Secondly, for a given direction, the total received power spectrum of the FB-based array is linearly proportional to the frequency response of the array's power gain, as described previously in *Section III*. Consequently, there is an easy and convenient approach to experimentally verify the proposed FB-enabled array technique—just use the configured setup in a microwave anechoic chamber for experimentation and the measured frequency response of the realized gain at a certain angle can be used to represent the practically received power spectrum from the FB-enabled array. The fabricated prototype of the proposed array is shown in Fig. 12(b), while the measurement setup is depicted in Fig. 12(c). Its measured $|S_{11}|$ and radiation patterns are plotted in Fig. 14 (a) and (b), respectively. These measured results are in a reasonably good agreement with their simulated counterparts, and both demonstrate the desirable spatial-stitching phenomenon. The realized gain frequency responses at several angles, which are used to represent the received power spectrums toward the relevant AoTs, are plotted in Fig. 14(c). Limited to the *Ka*-band measurement setup, only the measured results within 30-40 GHz are recorded and presented. The measured results of realized gain frequency responses are observed to reasonably agree with the simulated received power spectrums, thereby justifying the effectiveness of this simplified experimentation as well as the spectral-stitching effects. The measured 2-D realized gain pattern of the proposed array is plotted in Fig. 14(d), exhibiting an obvious expanded and near-doubled frequency-space area when compared to Fig. 10(c).

## V. Discussion and Conclusion

In this work, an enabling technique inspired by the FB concept and embodied in an array of LWAs has been proposed and investigated for potential FSR applications with full-band-locked beam-target illumination capabilities and decoupled range and angle resolutions. The proposed solution makes a combination of the natural frequency-based beam-scanning inherently owned by LWAs and the FB concept borrowed from the multi-rate signal processing field to realize a stitched frequency-space domain for radar operations. FB-related magnitude and phase stitching conditions in the form of the BSFs, beam-crossovers, and phase alignments have been theoretically derived for instructing the design of such an FB-enabled LWA array. Also, a detailed and generalized three-step design flow has been developed to facilitate practical implementations. However, there are several points that should be highlighted here. Firstly, our proposed solution mainly lies in the delicate engineering of an array of LWA channels in accordance with several FB-related specifications. The main body of a practical system architecture of our proposed solution can be almost the same as that of the traditional multi-channel FMCW radar sensor product [27] except for the antenna front-end and signal-processing back-end module. Secondly, the success or failure of realizing the required stitched frequency-space domain for radar signal processing is directly under the premise of whether or not the LWA array can be properly engineered or synthesized as designated. More specifically, if a stitched frequency-space coverage could be realized in the back-end module for radar signal processing, the whole FB-enabled LWA array itself must simultaneously present a widened radiation pattern ("spatial-stitching") at given frequencies and enhanced gain bandwidth ("spectral-stitching") towards given directions. This can be equivalently perceived as interpreting the "frequency-space stitching" phenomenon from the perspective of the antenna, which may give us a convenient and straightforward way to examine the effectiveness and correctness of the FB-enabled LWA array engineering prior to setting up such a relevant radar system. Under this context, a two-channel SLR-based microstrip combline LWA array has been modeled, simulated, and measured for a direct proof of concept, and the highly desired "frequency-space stitching" has been successfully demonstrated by both the simulated and measured results, thereby justifying our proposed antenna front-end solution. Notably, compared to the prevailing phased array technique [27][29]-[33], the virtues of our proposed FB-based LWA array scheme are distinct: it is shifter-less and full-band-locked (wideband), and the angle and range resolutions can be designed with a great degree of freedom for radar detection and sensing. For example, more priority can be given to the angle resolution without using plenty of Rx channels, which cannot be met by phased-array-based solutions [27][29]-[33]. As a result, the proposed scheme may be a potential candidate solution to be deployed in current and future radar applications [46].